\documentclass[prd, showpacs,preprintnumbers,nofootinbib]{revtex4}
\usepackage[dvips]{graphicx}
\usepackage{enumerate}
\usepackage{amsmath,amssymb}
\usepackage{mathrsfs}
\usepackage[dvips]{graphicx}
\usepackage{bm}

\begin{document}
\preprint{CHIBA-EP-206v2/KEK Preprint 2014-4, 2014}

\title{
Gauge-independent ``Abelian'' and magnetic-monopole dominance, 
\\
and the dual Meissner effect
in lattice $SU(2)$ Yang-Mills theory
}

\author{Seikou Kato$^{1}$}
\email{skato@fukui-nct.ac.jp}

\author{Kei-Ichi Kondo$^{2}$}
\email{kondok@faculty.chiba-u.jp}

\author{Akihiro Shibata$^{3}$}
\email{akihiro.shibata@kek.jp}


\affiliation{
$^1$Fukui National College of Technology, Sabae 916-8507, Japan
\\
$^2$Department of Physics,  
Graduate School of Science, 
Chiba University, Chiba 263-8522, Japan
\\
$^3$Computing Research Center,  High Energy Accelerator Research Organization (KEK),  
and Graduate University for Advanced Studies (Sokendai), Tsukuba  305-0801, Japan
}

\begin{abstract}
In the $SU(2)$ Yang-Mills theory on the four-dimensional Euclidean lattice, we confirm the gauge-independent ``Abelian'' dominance (or the restricted field dominance) and gauge-independent magnetic-monopole dominance in the string tension of the linear potential extracted from the Wilson loop in the fundamental representation. 
The dual Meissner effect is observed by demonstrating the squeezing of the chromoelectric field flux connecting a pair of quark and antiquark.
In addition, the circular magnetic-monopole current is induced around the chromoelectric flux.   
The type of the dual superconductivity is also determined by fitting the result with the dual Ginzburg-Landau model. 
Thus the dual superconductor picture for quark confinement is supported in a gauge-independent manner. 
These results are obtained based on a reformulation of the lattice Yang-Mills theory based on the change of variables a la Cho-Duan-Ge-Faddeev-Niemi combined with a non-Abelian Stokes theorem for the Wilson loop operator.  We give a new procedure (called the reduction) for obtaining the color direction field which plays the central role in this reformulation.

\end{abstract}

\pacs{12.38.Aw, 21.65.Qr}

\maketitle

\section{Introduction}

It is yet challenging to elucidate the  mechanism of color confinement in Quantum Chromodynamics (QCD).
The dual superconductor picture \cite{dualsuper} of the QCD vacuum may be a promising scenario for the  confinement mechanism. In particular, it is known that the string tensions calculated from the Abelian part and especially the magnetic monopole part reproduce well the original string tension, once we perform an Abelian projection \cite{tHooft81} in  the Maximally Abelian (MA) gauge \cite{KLSW87}. 
These phenomena are respectively called the  Abelian  dominance \cite{SY90,AS99} and magnetic monopole dominance \cite{SNW94}.
The magnetic monopole in the Abelian projection is the gauge fixing defect defined using the partial gauge fixing from the original non-Abelian gauge group $G$ to the maximal torus subgroup $H$. 
Therefore, we cannot judge whether the results obtained in the MA gauge are gauge independent physical phenomena or not. 
In fact, these phenomena can be seen only in specific gauges, i.e., Abelian gauges, such as the MA gauge,  Laplacian Abelian gauge and maximal center gauge \cite{Greensite03}.
See, e.g., \cite{CP97} for reviews of Abelian projection and Abelian gauge.

Recently, we have given a reformulation of the Yang-Mills theory based on a new viewpoint proposed in \cite{KMS05,KMS06,Kondo06} using the new field variables obtained through the change of variables from the original Yang-Mills field, which originates from the field decomposition called the  Cho--Duan-Ge (CDG) or Cho--Duan-Ge--Faddeev--Niemi (CDGFN) decomposition 
\cite{Cho80,DG79,FN99,Shabanov99}. 
The reformulation enables us to define a manifestly gauge-invariant magnetic monopole in the pure Yang-Mills theory without any fundamental scalar field, which does not need the Abelian projection. 
We have constructed the lattice version of the reformulation to perform numerical simulations. 
We have found that the magnetic charge of the resulting  magnetic monopole is perfectly quantized on the lattice \cite{KKMSSI06}.
Moreover, we have confirmed that the resulting magnetic monopole dominates the string tension \cite{Kato-lattice2009,IKKMSS06}, 
although  the magnetic monopole dominance in the string tension  was first shown using the conventional MA gauge in \cite{SNW94}. 
Therefore we suggest the gauge independence of the dual superconductor picture of the QCD vacuum.
The infrared Abelian dominance was confirmed also for the correlation functions on the lattice \cite{SKKMSI07}. 
These results are obtained based on a non-Abelian Stokes theorem for the Wilson loop operator \cite{DP89,Kondo98b,KT99,Kondo99Lattice99,Kondo08}.  
We give a new procedure (called the reduction) for obtaining the color direction field which plays the central role in this reformulation.

In this paper, we give more numerical evidences for the dual superconductor picture of the QCD vacuum.
We give more data supporting the gauge-independent infrared  ``Abelian'' dominance (or the restricted field dominance) and gauge-independent magnetic monopole dominance in the string tension of the linear potential, which is derived from the calculation of the Wilson loop average on a larger lattice. 
The dual Meissner effect is observed by demonstrating the squeezing of the chromoelectric field flux connecting a pair of quark and antiquark.
In addition, the circular magnetic-monopole current is induced around the chromoelectric flux.   
The type of the dual superconductivity is also determined by fitting the result with the dual Ginzburg-Landau model. 
Thus the dual superconductor picture for quark confinement is established in a gauge-independent manner. 
In addition, we give a new procedure (called the reduction) for obtaining the color direction field which plays the central role in the reformulation.

We focus our investigations on the $SU(2)$ gauge group,
although the formulation has been extended to $SU(N)$ gauge group in the continuum \cite{Cho80c,FN99a,KSM08}  and  on the lattice \cite{KSSMKI08,SKS10,KS08}. 
For $SU(3)$, the results of numerical simulations are given in  \cite{SKKS11,SKKS12}. 
See \cite{KKSS14} for a review, and \cite{KKSS14a} for the digest of the analytical results and \cite{SKKS14b} for the numerical results.

\section{Reformulation of  lattice SU(2) Yang-Mills theory}

In the preceding papers \cite{KKMSSI06,IKKMSS06,SKKMSI07,KSSMKI08,SKS10}, we have proposed a novel reformulation of  lattice Yang-Mills theory written in terms of new variables, which is obtained by   change of variables from the ordinary lattice Yang-Mills field.
For the $SU(2)$ gauge group, the lattice reformulation reproduces  the Cho-Duan-Ge-Faddeev-Niemi decomposition for $SU(2)$ case \cite{Cho80,DG79,FN99,Shabanov99} in the naive continuum limit, while for the $SU(N)$ gauge group $(N \ge 3)$, it reproduces the continuum counterpart proposed in \cite{KSM08}, which includes the   Cho-Faddeev-Niemi decomposition for $SU(N)$ case \cite{Cho80c,FN99a}   as a special case.

In this paper, we focus our studies on the $SU(2)$ case on a lattice, since the $SU(3)$ case are studied in \cite{SKKS11,SKKS12,KS08}.
Let $\mathscr{A}_\mu(x)$ be the Yang-Mills field  taking the values in the Lie algebra $su(2)$ of the $SU(2)$ group: 
\begin{align}
 \mathscr{A}_\mu(x) = \mathscr{A}_\mu^A(x)T^A \in su(2), \quad T^A=\frac12 \sigma_A \ (A=1,2,3) , 
\label{def-A}
\end{align}
where $\sigma_A$ ($A=1,2,3$) are the Pauli matrices. 
 On a lattice with a lattice spacing $\epsilon$,  a gauge variable taking the values in the   gauge group $G$ is represented as a link variable $U_{x,\mu}$ defined on a link $<x,x+ \epsilon \hat\mu>$, 
\begin{align}
U_{x,\mu} = \exp( -i \epsilon g \mathscr{A}_\mu(x^\prime)) \in G=SU(2) , 
\label{def-U1}
\end{align}
where $x^\prime$ is the midpoint $x^\prime:=x+ \epsilon \hat\mu/2$ of the link $<x,x+ \epsilon \hat\mu>$.
Here $g$ denotes the coupling constant.
The midpoint prescription is useful to suppress as much as possible lattice artifacts coming from a finite (nonzero) lattice spacing, in contrast to the very naive definition  
between the gauge link variable $U_{x,\mu}$ and the gauge potential $\mathscr{A}_\mu(x)$  given by 
\begin{align}
U_{x,\mu} = \exp( -i \epsilon g \mathscr{A}_\mu(x)) \in SU(2)  . 
\label{def-U}
\end{align}


The link variable $U_{x,\mu}$ obeys the well-known lattice gauge transformation:
\begin{equation}
  U_{x,\mu}  \rightarrow U_{x,\mu}^\prime = \Omega_{x} U_{x,\mu} \Omega_{x+\mu}^{\dagger} 
  , \quad \Omega_{x} \in SU(2)
  \label{U-transf}
 .
\end{equation}

In order to define new variables, we consider the decomposition of the $SU(2)$-valued gauge variable $U_\ell=U_{x,\mu}$ into the product of two $SU(2)$-valued variables $X_{x,\mu}$ and $V_{x,\mu}$ defined on the same lattice \cite{KSSMKI08}:
\footnote{
Here the lattice variables $V_{x,\mu}$ and $X_{x,\mu}$ are supposed to be related to the Lie-algebra $\mathscr{V}_\mu(x)$ and  $\mathscr{X}_\mu(x)$ in the continuum as  
\begin{equation}
  V_{x,\mu} = \exp \{-i\epsilon g\mathscr{V}_\mu(x^\prime) \}  , \ 
  X_{x,\mu} = \exp \{-i\epsilon g\mathscr{X}_\mu(x) \}   
  ,
\end{equation}
just as
$
  U_{x,\mu} = \exp \{ -i\epsilon g\mathscr{A}_\mu(x^\prime) \} 
$.
However, the decomposition can be constructed so as to have an intrinsic meaning on a lattice without referring to the naive continuum limit \cite{SKS10}.
}
\begin{align}
SU(2) \ni U_{x,\mu} = X_{x,\mu}V_{x,\mu}  , \quad X_{x,\mu}  \in  SU(2)  , \ V_{x,\mu} \in  SU(2)  ,
\label{NLCV-1}
\end{align}
in such a way that 
  $V_{x,\mu}$ transforms just like a usual link variable under the gauge transformation:
\begin{align}
  V_{x,\mu} \rightarrow V_{x,\mu}' = \Omega_{x} V_{x,\mu} \Omega_{x+\mu}^\dagger    , \quad  \Omega_{x} \in  SU(2)  
  \label{gt-V}
\end{align}
and thereby $X_{x,\mu}$ transforms like a site  variable under the gauge transformation:
\begin{align}
  X_{x,\mu} \rightarrow X_{x,\mu}' = \Omega_{x} X_{x,\mu} \Omega_{x}^\dagger  , \quad  \Omega_{x} \in  SU(2)  .
  \label{gt-X}
\end{align}

Such a decomposition can be performed by introducing another field: 
In the reformulation, we introduce the site variable taking the values in the  Lie algebra  of $SU(2)/U(1)$:
\begin{align}
  \bm{n}_x  :=n_x^A \sigma_A \in Lie(SU(2)/U(1)) = su(2)-u(1) .
\end{align}
We call $\bm{n}_x$ a {{{color (direction)  field}, since the color field $\bm{n}_{x}$ is used to specify only the color direction in the color space at each spacetime point and its magnitude is irrelevant ($\bm{n}_{x}^2=\bm{n}_{x} \cdot \bm{n}_{x} =1$).
It should be remarked that the color field $\bm{n}_{x}$ is Hermitian,  $\bm{n}_{x}^\dagger=\bm{n}_{x}$, 
while the gauge field $U_{x,\mu}$ is unitary, $U_{x,\mu}^\dagger=U_{x,\mu}^{-1}$.   

The reformulation is constructed such that the site variable ${\bf n}_{x}$ transforms under the gauge transformation (which was called the gauge transformation II in \cite{KMS05}) as
\begin{align}
  \bm{n}_{x} \rightarrow \bm{n}_{x}^\prime = \Omega_{x} \bm{n}_{x} \Omega_{x}^{\dagger}   , \quad \Omega_{x}  \in SU(2)  . 
  \label{gt-n}
\end{align}

It is shown that the decomposition is uniquely determined by imposing the two requirements called the {{{ defining equation}:%
\begin{enumerate}
\item[(i)]
 the color field $\bm{n}_{x}$ is covariantly 
constant in the background (matrix) field $V_{x,\mu}$:
\begin{align}
 \bm{n}_{x} V_{x,\mu}  = V_{x,\mu} \bm{n}_{x+\mu} 
\Longleftrightarrow 
D_\mu^{(\epsilon)}[\mathscr{V}] \bm{n}_{x} = 0 ,
 \label{Lcc}
\end{align}

\item[(ii)]
 the remaining (matrix) field $X_{x,\mu}$ is perpendicular to the color field  $\bm{n}_{x}$:%
\footnote{
This requirement can be replaced by the exact form in the compact formulation, see \cite{SKS10}. 
}
\begin{equation}
 {\rm tr}(\bm{n}_{x} X_{x,\mu} ) \equiv {\rm tr}(\bm{n}_{x} U_{x,\mu} V_{x,\mu}^\dagger) 
  = 0 .
  \label{cond2m}
\end{equation}
\end{enumerate}
 Both conditions (i) and (ii) must be imposed to uniquely determine $V_{x,\mu}$ and $X_{x,\mu}=U_{x,\mu}V_{x,\mu}^{-1}$ for a given set of 
$U_{x,\mu}$ once the color field $\bm{n}_{x}$ is determined. 
They are the naive lattice version of the defining equations in the continuum.
In the naive continuum limit $\epsilon \rightarrow 0$, indeed, these defining equations reduce to the continuum counterparts.
It is important to remark that these defining equations are covariant or form-invariant under the gauge transformation, which is necessary for  the decomposed variables to have the desired transformation property (\ref{gt-V}), (\ref{gt-X}) and (\ref{gt-n}). 
In fact, the  defining equation (\ref{Lcc})   is form-invariant under the gauge transformation (\ref{gt-n}) and (\ref{gt-V}), i.e.,
$ \bm{n}_{x}^\prime V_{x,\mu}^\prime  = V_{x,\mu}^\prime \bm{n}_{x+\mu}^\prime
$. 
This is also the case for the second defining equation (\ref{cond2m}):
$
 {\rm tr}(\bm{n}_{x}^\prime X_{x,\mu}^\prime ) =0 
$.

We can solve the defining equation (\ref{Lcc}) for the link variable $V_{x,\mu}$ and express it in terms of the site variable $\bm{n}_{x}$ and the original link variable $U_{x,\mu}$, 
just as the continuum variable $\mathscr{V}_\mu(x)$ is expressed in terms of $\bm{n}(x)$ and $\mathscr{A}_\mu(x)$. 
By solving the  defining equation (\ref{Lcc}) and (\ref{cond2m}), indeed, 
the link variable $V_{x,\mu}$ is determined  up to an overall normalization constant 
in terms of the site variable $\bm{n}_{x}$ and the original link variable 
$U_{x,\mu}$ \cite{IKKMSS06}:%
\begin{align}
  \tilde{V}_{x,\mu} = \tilde{V}_{x,\mu}[U,\bm{n}] 
  = U_{x,\mu} +   \bm{n}_{x} U_{x,\mu} \bm{n}_{x+\mu} .
  \label{sol}
\end{align}
The equation (\ref{Lcc}) is linear in $V_{x,\mu}$. Therefore, the normalization of $V_{x,\mu}$ cannot be determined by this equation alone.
In general,  unitarity is not guaranteed for the general solution of the defining equation and hence a unitarity condition must be imposed afterwards. 
Fortunately, this issue is easily solved at least for $SU(2)$ group,  
since the speciality condition $\det V_{x,\mu}=1$ determines the normalization. 
Then the special unitary link variable $V_{x,\mu}[U,\bm{n}]$ is obtained after the normalization: 
\begin{align} 
V_{x,\mu} = 
V_{x,\mu}[U,\bm{n}] := 
 \tilde{V}_{x,\mu}/\sqrt{\frac{1}{2}{\rm tr} [\tilde{V}_{x,\mu}^{\dagger}\tilde{V}_{x,\mu}]} .
\label{cfn-mono-4}
\end{align}
It is  shown \cite{IKKMSS06} that the naive continuum limit $\epsilon \rightarrow 0$ 
of the link variable $V_{x,\mu} = \exp (-i\epsilon g \mathscr{V}_\mu(x))$ reduces to the continuum expression:
\begin{align} 
\mathscr{V}_{\mu}(x) = (n^A(x)A_{\mu}^A(x))\bm{n}(x)
-ig^{-1} [\partial_{\mu}\bm{n}(x) ,  \bm{n}(x) ] ,
\label{cfn-conti-7} 
\end{align}
which  agrees with the expression of the restricted field in the Cho-Duan-Ge decomposition  in the continuum \cite{Cho80,DG79}. 
This is indeed the case for the remaining variable $X_{x,\mu} = \exp (-i\epsilon g  \mathscr{X}_\mu(x))$.

 By including the color field $\bm{n}_x$, the $SU(2)$ Yang-Mills theory written in terms of $U_{x,\mu}$ is extended to a gauge theory written in terms of $U_{x,\mu}$ and $\bm{n}_x$ with the enlarged local gauge symmetry $\tilde{G}_{\omega,\theta}^{local}=SU(2)^{local}_{\omega} \times [SU(2)/U(1)]^{local}_{\theta}$ 
larger than the local gauge symmetry $SU(2)^{local}_{\omega}$ in the original 
Yang-Mills theory \cite{KMS05}. 
  In order to eliminate the extra degrees of freedom in the enlarged local gauge symmetry 
$\tilde{G}_{\omega,\theta}^{local}$ for obtaining the Yang-Mills theory which is equipollent to the original Yang-Mills theory, 
we must impose sufficient number of constraints, which we called the {{{reduction condition}.

We find that such a  reduction condition is given on a lattice by minimizing the functional:
\begin{align}
F_{\rm red}[\bm{n};U] &=  \sum_{x,\mu} \left[ 1- 
                   {\rm tr}(\bm{n}_xU_{x,\mu}\bm{n}_{x+\hat{\mu}}U_{x,\mu}^{\dagger})
                   / {\rm tr}({\bf 1}) \right] ,
\end{align}
with respect to the color  fields $\{\bm{n}_x\}$ for a given set of link variables $\{U_{x,\mu}\}$.
Thus color field $\bm{n}_x$ is determined by $\bm{n}_x=\bm{n}_x^*$ in such a way that the functional achieves the minimum at $\bm{n}_x=\bm{n}_x^*$:
\begin{align}
{\rm min}_{\bm{n}} F_{\rm red}[\bm{n};U] = F_{\rm red}[\bm{n}^*; U].
\label{reduction_n}
\end{align}

The two algorithms for solving the reduction equation are available:
\begin{enumerate}
\item[(i)]
Updating $\{ \bm{n}_x \}$ via gauge transformation for  solving the reduction condition.
 This method  of the reduction prescription  was adopted in the early studies \cite{IKKMSS06,SKKMSI07}. (This was once called the new MAG.) 

\item[(ii)]
 We solve the stationary condition:
\begin{align}
\frac{\partial F_{\rm red}[\bm{n};U]}{\partial n^A_x} =0 ,
\end{align}
in order to minimize the functional $F_{\rm red}$.
This method of the reduction prescription was adopted in this paper.
\end{enumerate}
The functional $F_{\rm red}$ can be rewritten in the following way:
\begin{align}
F_{\rm red}[\bm{n};U] 
=&  \sum_{<x,y>} (1-J^{AB}_{x,y}[U]n^A_xn^B_y)  
  , 
  \quad
J^{AB}_{x,y}[U] =   {\rm tr}(\sigma^A U_{x,\mu}\sigma^B U_{x,\mu}^{\dagger})/{\rm tr}({\bf 1}) .
\end{align}
Therefore, the functional $F_{\rm red}$ can be regarded as the energy for the spin-glass system.

There exist local minima which satisfy the reduction condition. 
Therefore, overrelaxation method
should be used in order to approach the global minimum more rapidly.

\section{Gauge-independent ``Abelian'' dominance and magnetic-monopole dominance in the Wilson loop}

The Wilson loop operator $W_{\rm full}[U]$ for a closed loop $C$ on a lattice is defined using the link variable  $U_\ell$ in the gauge-invariant way:
\begin{align}
W_{\rm full}[U] := {\rm tr}({\cal P}\prod_{\ell \in C}U_\ell)/{\rm tr}({\bf 1})  .
\end{align}
By replacing the  full $SU(2)$  link variable $U_\ell$ by the restricted variable $V_\ell$, we can define another gauge-invariant quantity $W_{\rm rest}[V]$ which we call the {{{restricted Wilson loop operator}:
\begin{align}
W_{\rm rest}[V] := {\rm tr}({\cal P}\prod_{\ell \in C}V_\ell)/{\rm tr}({\bf 1}) . 
\end{align}
Then we can define the Wilson loop average $W_{\rm full}(C)$ and the restricted Wilson loop average $W_{\rm rest}(C)$ by
\begin{align}
W_{\rm full}(C) := \left< W_{\rm full}[U] \right> , \quad  
W_{\rm rest}(C) := \left< W_{\rm rest}[V] \right> .
\end{align}
Therefore, the respective average  must be independent of the gauge.
Since the restricted field ${\bf V}_{\mu}(x)$ is defined in a gauge-covariant and gauge independent way, 
we have obtained a  gauge-independent definition of the {{{``Abelian'' 
dominance} or the {{{restricted-field dominance} for the Wilson loop average:
\begin{align}
  W_{\rm full}(C)  \simeq  {\rm const.} W_{\rm rest}(C) .
\end{align}
In \cite{KS08},  a gauge-independent definition of Abelian dominance was given in the operator level  $ W_{\rm full}[U]  \simeq   {\rm const.}   W_{\rm rest}[V]$ and a constructive derivation of the Abelian dominance was discussed  through a non-Abelian Stokes theorem via lattice regularization.


In the reformulation, moreover, we can define the {{{gauge-invariant field strength} $\bar{\Theta}_P[U,{\bf n}]$ 
 as a plaquette variable on a lattice by \cite{IKKMSS06}
\begin{align} 
\bar{\Theta}_{x,\mu\nu}[U,\bm{n}] := \epsilon^{-2}
{\rm arg} [ {\rm tr} \{({\bf 1}+\bm{n}_x)V_{x,\mu}V_{x+\hat{\mu},\nu}
V_{x+\nu,\mu}^{\dagger}V_{x,\nu}^{\dagger} \}/{\rm tr}({\bf 1})] .
\label{cfn-mono-5}
\end{align}
Taking into account the relation:
\begin{align} 
 V_{P} =  V_{x,\mu}V_{x+\hat{\mu},\nu}
V_{x+\nu,\mu}^{\dagger}V_{x,\nu}^{\dagger} 
&= \exp \{ -i\epsilon^2 g \mathscr{F}_{\mu\nu}[\mathscr{V}] \}
, \quad
 \mathscr{F}_{\mu\nu}[\mathscr{V}]  := \partial_\mu \mathscr{V}_{\nu} - \partial_\nu \mathscr{V}_{\mu} -i g [ \mathscr{V}_{\mu} ,  \mathscr{V}_{\nu} ] ,
\end{align}
and the expansion:
\begin{align} 
& V_{P} =    {\bf 1} -i\epsilon^2 g \mathscr{F}_{\mu\nu}[\mathscr{V}] + O(\epsilon^4) ,
 \quad 
 {\rm tr} ( V_{P} ) =  {\rm tr} ({\bf 1} ) + O(\epsilon^4) ,
\nonumber\\
& {\rm tr} ( \bm{n}_x V_{P} )  
=   -i\epsilon^2 g {\rm tr} ( \mathscr{F}_{\mu\nu}[\mathscr{V}] \bm{n}_x ) + O(\epsilon^4) 
=  - i\epsilon^2 g \frac12  \mathscr{F}_{\mu\nu}[\mathscr{V}] \cdot \bm{n}_x   + O(\epsilon^4)
 ,
\end{align}
it is shown  that the naive continuum limit of (\ref{cfn-mono-5}) reduces to
the gauge-invariant field strength (see Appendix A.2 of \cite{IKKMSS06}):
\begin{align} 
\bar{\Theta}_{x,\mu\nu} \simeq&
\partial_{\mu}(n^A(x) \mathscr{A}_{\nu}^A(x))-\partial_{\nu}(n^A(x) \mathscr{A}_{\mu}^A(x))
-i g^{-1} \bm{n}\cdot [\partial_{\mu}\bm{n}, \partial_{\nu}\bm{n} ]
\nonumber\\
=&\frac{-1}{2} \bm{n} \cdot \mathscr{F}_{\mu\nu}[\mathscr{V}]  .
\label{cfn-fs} 
\end{align}
Here, $\bar{\Theta}_{x,\mu\nu}$  plays the similar role to the 't Hooft  tensor in describing the 
't Hooft--Polyakov magnetic monopole in Georgi--Glashow model.
\footnote{
The lattice definition which reduces to the the continuum form $ \bm{n} \cdot \mathscr{F}_{\mu\nu}[\mathscr{V}]$ in the naive continuum limit is not unique, e.g., 
${\rm tr} \{ {\bf 1}+ \bm{n}_x V_{P} \}$ has the same form as 
${\rm tr} \{({\bf 1}+ \bm{n}_x)V_{P} \}$
up to $O(\epsilon^2)$. 
The advantage of using the form ${\rm tr} \{({\bf 1}+ \bm{n}_x)V_{P} \}$ is that it guarantees the quantization of the magnetic charge. 
}

Then we can define the {{{gauge-invariant magnetic-monopole current}. 
We use the gauge-invariant field strength $\bar{\Theta}_{x,\mu\nu}[V,{\bf n}]$ to extract  configurations of 
the magnetic-monopole current $\{K_{x,\mu}\}$  defined by the integer-valued field $m_{x,\mu}$:
\begin{align}
K_{x,\mu}= 2\pi m_{x,\mu}, \quad 
m_{x,\mu} = 
-\frac{1}{4\pi}{\varepsilon}_{\mu\nu\rho\sigma}
\partial_{\nu}\bar{\Theta}_{x+\mu,\rho\sigma} \in \mathbb{Z} .
\label{cfn-conti-20}
\end{align}
This definition satisfies the quantization of the magnetic charge \cite{IKKMSS06}. 
This definition of the magnetic-monopole  current $\{K_{x,\mu}\}$ agrees with our definition of the magnetic-monopole current in the continuum limit (divided by $2\pi$).

In order to study the {{{magnetic-monopole dominance}  in the string tension, we proceed to estimate the 
magnetic monopole contribution: 
\begin{align}
W_{\rm mono}(C)=\left< W_{\rm mono}[K] \right> 
\end{align}
to the Wilson loop average $ W_{\rm full}(C)=\left< W_{\rm full}[U] \right> $.  
Here we define the magnetic part $W_{\rm mono}[K]$ of the Wilson loop operator $W_{\rm full}[U]$ as the contribution 
from the monopole current $K_{x,\mu}$ to the Wilson loop operator:
\begin{align}
 W_{\rm mono}[K]     &:=  \exp \left( i \epsilon^4 \sum_{x,\mu} K_{x,\mu}\Xi_{x,\mu} \right) 
= \exp \left( 2\pi i \epsilon^4 \sum_{x,\mu}m_{x,\mu}\Xi_{x,\mu} \right) ,
\nonumber\\
 \Xi_{x,\mu}  &:=  \epsilon^4 \sum_{x'}\Delta_L^{-1}(x-x')\frac{1}{2}
\epsilon_{\mu\alpha\beta\gamma}\partial_{\alpha}
S^J_{x'+\epsilon \hat{\mu},\beta\gamma}, 
\quad
\partial'_{\beta}S^J_{x,\beta\gamma} = J_{x,\gamma} ,
\label{monopole dominance-2}
\end{align}
where $\Xi_{x,\mu}$ is defined through the external source $J_{x,\mu}$ which is used 
to calculate the static potential, 
$\partial'$ denotes the backward lattice derivative
$\partial_{\mu}^{'}f_x :=\epsilon^{-1}(f_x-f_{x-\mu})$,  $S^J_{x,\beta\gamma}$ denotes a surface bounded by the closed loop $C$ on which the electric source $J_{x,\mu}$ has its support, and $\Delta_L^{-1}(x-x')$ is the inverse Lattice Laplacian. 
Note that $W_{\rm mono}[K]$ is a gauge-invariant operator, since the monopole current defined
by (\ref{cfn-conti-20}) is a gauge-invariant variable.
In fact, the form (\ref{monopole dominance-2}) is derived from the non-Abelian Stokes theorem for the Wilson loop operator. 
By evaluating the average of $W_{\rm mono}[K]$ from the generated configurations of the monopoles $\{ K_{x,\mu}\}$ we can estimate the contribution to the string tension  from the generated configurations of the magnetic monopole currents $\{K_{x,\mu}\}$.

The Wilson loop operator $W_{\rm full}[U]$ is decomposed into the magnetic part $W_{\rm mono}[K] $ and the electric part $W_{\rm elec}[j] $, 
\begin{align}
 W_{\rm full}[U] = W_{\rm mono}[K] W_{\rm elec}[j],
\end{align}
which is derived from the non-Abelian Stokes theorem, see  \cite{Kondo08}.
The magnetic part $W_{\rm mono}[K]$ of the Wilson loop operator $W_{\rm full}[U]$ is used to examine the contribution from the monopole current $k_{x,\mu}$ to the Wilson loop operator $W_{\rm full}[U]$, while $W_{\rm elec}[j]$ is expressed by the electric current $j_{\mu}=\partial_{\nu}F_{\mu\nu}$.
In order to establish the monopole dominance in the string tension, we proceed to estimate the magnetic monopole contribution $\left< W_{\rm mono}[K] \right>$ to the Wilson loop average $\left< W_{\rm full}[U] \right>$, i.e., the expectation value of the Wilson loop operator.%
It should be remarked that $\left< W_{\rm full}[U] \right> \not= \left< W_{\rm mono}[K] \right> \left< W_{\rm elec}[j] \right>$.
We have not yet calculated the electric contribution $\left< W_{\rm elec}[j]  \right>$ directly  where 
$W_{\rm elec}[j] $ is expressed by the electric current $j_{\mu}=\partial_{\nu}F_{\mu\nu}$. 
See \cite{Kondo08} for details. 

\section{Numerical simulations of $SU(2)$ Yang-Mills theory}

In what follows, we present the results of numerical simulations. 
First of all, we generate the configurations of $SU(2)$ link variables 
$\{ U_{x,\mu} \}$, 
using the (pseudo) heat bath method for the standard Wilson action. 

Second, we generate the configurations of the color field $\{{\bf n}_x\}$ using the reduction condition (\ref{reduction_n}) for the obtained configurations of $SU(2)$ link variables $\{U_{x,\mu}\}$. 
Then we can construct the restricted field $\{V_{x,\mu}[U,{\bf n}]\}$ according to the change of variables (\ref{cfn-mono-4}).
Moreover, we can construct the magnetic-monopole current $\{k_{x,\mu}\}$ according to (\ref{cfn-conti-20}).

\subsection{Wilson loop average and the quark potential}

For a rectangular Wilson loop $C=(R,T)$ with the spatial length $R$ and the temporal length $T$, we calculate the three kinds of the Wilson loop average $W_{\rm i}(C)$ (\text{i=f(full)}, \text{r(rest)}, \text{m(mono)}).
Then we calculate the static $q\bar q$ potential $V_{\rm i}(R)$ as a function of the interquark distance $R$ using the respective Wilson loop 
average $W_{\rm i}(C)$ according to 
\begin{align}
V_{\rm i}(R) = -\log \left\{ \frac{  W_{\rm i}(R,T) }{ W_{\rm i}(R,T-1) } \right\} 
\quad (\text{i=f(full), r(rest), m(mono)}) .
\label{monopole dominance-3}
\end{align}

The numerical simulations are performed on the   $16^4$ lattice at $\beta=2.4$  and $24^4$ lattice at $\beta=2.5$.%
\footnote{
The lattice spacing in the physical units is given by \cite{SKKMSI07} $\epsilon(\beta=2.4)=0.1201$fm, and 
$\epsilon(\beta=2.5)=0.08320$fm.
}
We thermalize 3000 sweeps, and in particular, we have used 100 configurations for calculating  the full potential $V_{\rm full}$ and restricted potential $V_{\rm rest}$,  while for the monopole potential $V_{\rm mono}$  we have used 50 configurations for the $16^4$ lattice and 500 configurations for the $24^4$ lattice in each case with 100 iterations.%
\footnote{
The results of numerical simulations on the $16^4$  lattice at $\beta=2.4$ were published in \cite{IKKMSS06} only for the full potential $V_{\rm full}$ and the monopole potential $V_{\rm mono}$, while the result on  the restricted potential $V_{\rm rest}$ was separately reported in \cite{Kato-lattice2009}.
} 
In order to obtain the full $SU(2)$ and restricted  results, especially, we used the  {{{smearing method} \cite{albanese87} as a noise reduction technique.

Fig.\ref{fig:potential} shows the obtained plot for the respective potential for various values of $R$.
The obtained numerical potential is fitted to the sum of a linear term,  Coulomb term and a constant term: 
\begin{align}
V_{\rm i}(R) = \sigma_{\rm i} R -  \alpha_{\rm i}/R +c_{\rm i} , 
\quad (\text{i=f(full), r(rest), m(mono)}) .
\label{monopole dominance-4}
\end{align}
where $\sigma_{\rm i}$ is the string tension (the coefficient of the area decay), $\alpha_{\rm i}$ is the Coulomb coefficient, 
and $c_{\rm i}$ is the constant which is equal to the coefficient of the perimeter decay:
\begin{equation} 
W_i(R,T)  \sim \exp [-\sigma_i RT -c_i(R+T)+\alpha_i T/R + \cdots] .
\end{equation}
The results are shown in 
 Table~\ref{strint-tension-1a},
 Table~\ref{strint-tension-1}, and Figure~\ref{fig:potential}.

%

Thus, on the $16^4$ lattice~at $\beta=2.4$  the restricted (``Abelian'') part $\sigma_{\rm rest}$ reproduces 93$\%$ of the full string tension $\sigma_{\rm full}$:
\begin{equation} 
 \frac{\sigma_{\rm rest}}{\sigma_{\rm full}} = (93 \pm 17)\% 
\quad (\text{on} \ 16^4 \ \text{lattice~at} \ \beta=2.4), 
\end{equation}
and the monopole part $\sigma_{\rm mono}$ reproduces 94$\%$ of $\sigma_{\rm rest}$: 
\begin{equation} 
 \frac{\sigma_{\rm mono}}{\sigma_{\rm rest}} =(94 \pm 8)\%  
 \Longrightarrow 
 \frac{\sigma_{\rm mono}}{\sigma_{\rm full}} = (88 \pm 13)\% 
\quad (\text{on} \ 16^4 \ \text{lattice~at} \ \beta=2.4) . 
\end{equation}

\begin{figure}[ptb]
\begin{center}
\includegraphics[scale=0.65]{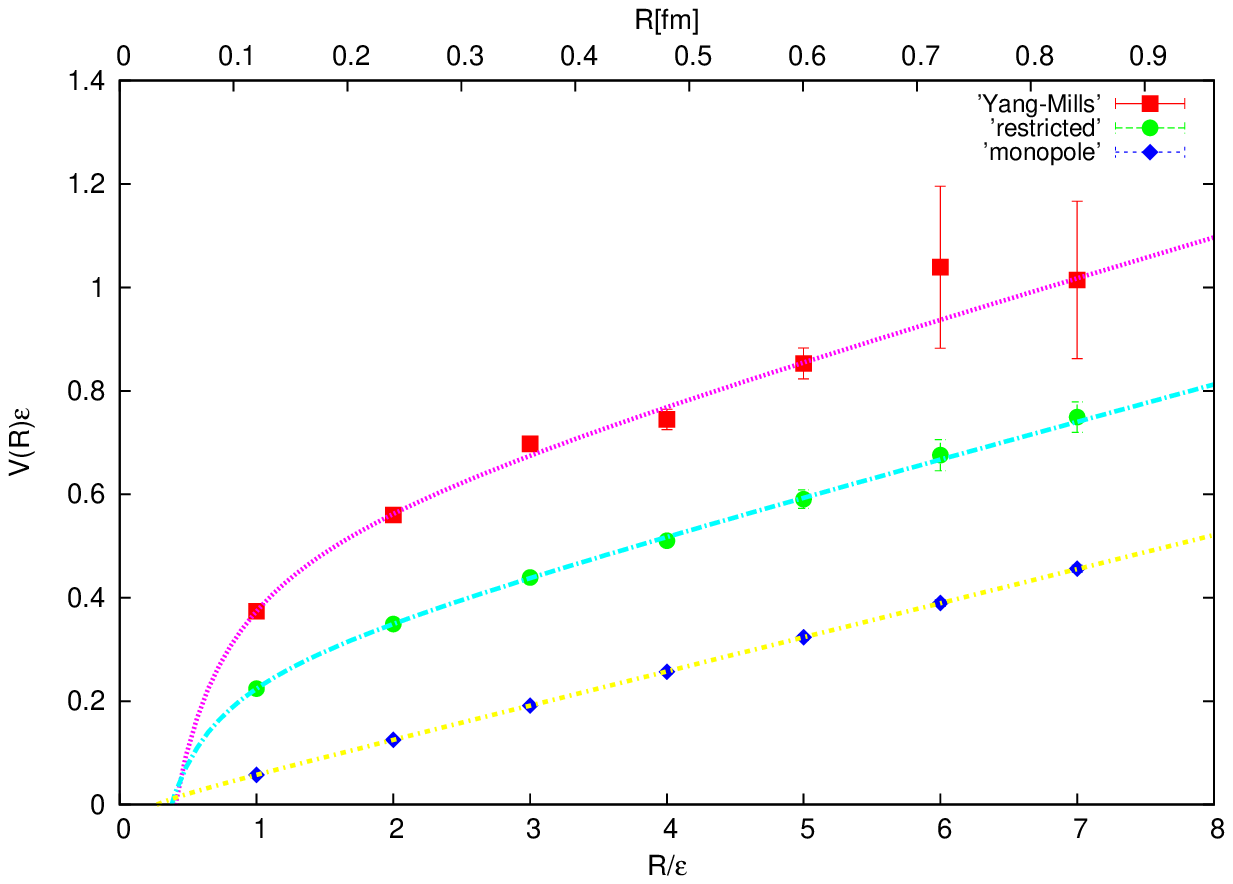} 
\quad 
\includegraphics[scale=0.65]{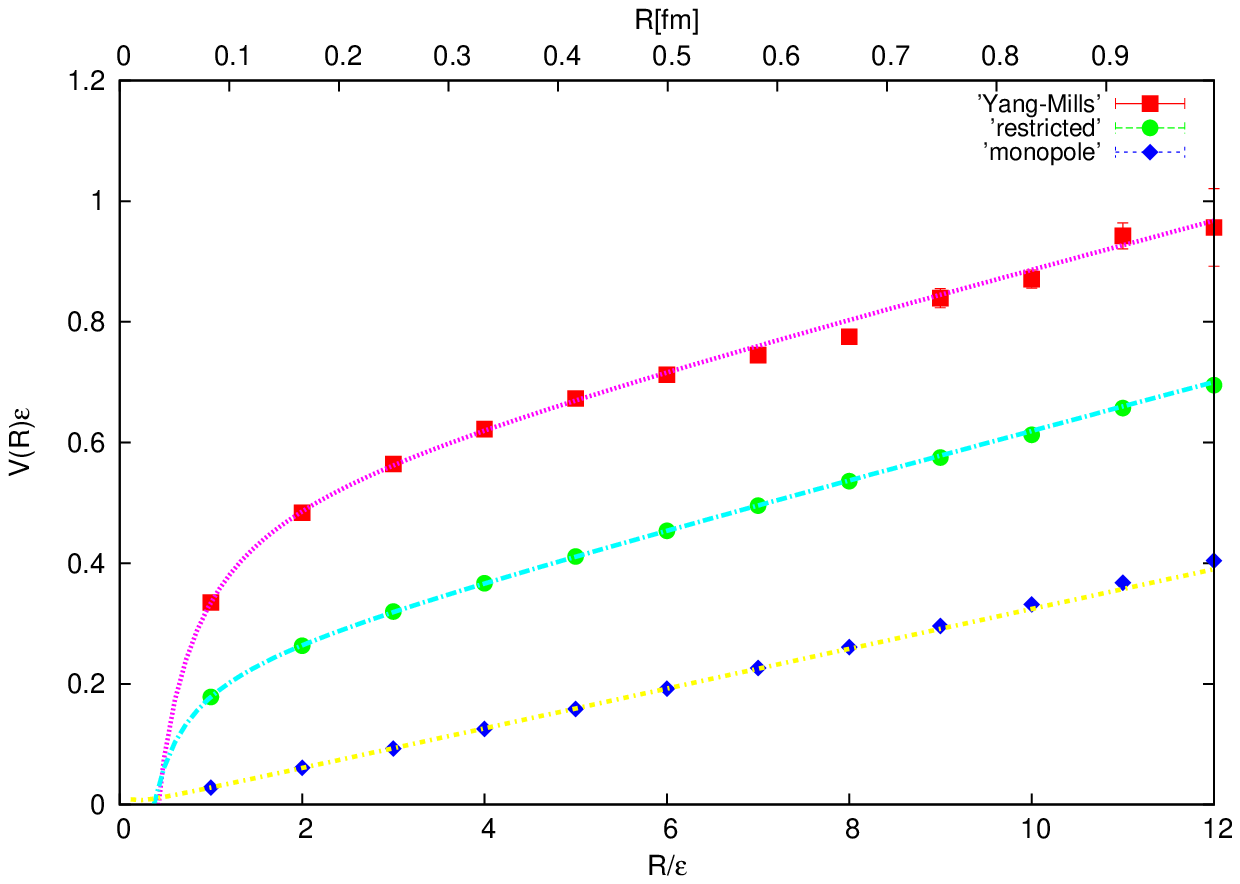}
\end{center}
\vspace{-5mm}
\caption{
The full $SU(2)$ potential $V_f(R)$,  (``Abelian'') restricted potential $V_r(R)$ and magnetic-monopole potential $V_m(R)$ as functions of $R$ 
(Left) on $16^4$ lattice at $\beta=2.4$,
(Right)  on $24^4$ lattice at $\beta=2.5$ 
where the Wilson loop with $T=12$ was used for obtaining  $V_{\rm full}(R)$ and $V_{\rm rest}(R)$, and $T=8$ for  $V_{\rm mono}(R)$. 
}%
\label{fig:potential}
\end{figure}

\begin{table}[ht]
\caption{String tension and Coulomb coefficient on $16^4$ lattice at $\beta=2.4$.}
\vspace{-3mm}
\begin{center}
\begin{tabular}{lllc}\hline
             & $\sigma$    & $\alpha$ & $\chi^2/N_{df}$\\ \hline
full     & 0.075(9)    & 0.23(2)  &   1.234 \\
restricted  & 0.070(4)    & 0.11(1) &  0.195 \\
magnetic monopole     & 0.066(2)    & 0.003(7) & 0.198\\
\hline
\end{tabular}
\end{center}
\label{strint-tension-1a}
\end{table}

\begin{table}[ht]
\caption{String tension and Coulomb coefficient on $24^4$ lattice at $\beta=2.5$.}
\vspace{-3mm}
\begin{center}
\begin{tabular}{lllc}\hline
             & $\sigma$    & $\alpha$  & $\chi^2/N_{df}$ \\ \hline
full      & 0.0388(6)    & 0.2245(23)  & 4.73 \\
restricted       & 0.0398(2)    & 0.0912(8)  & 1.82 \\
magnetic monopole    & 0.0330(1)    & -0.0012(4) & 4.81 \\
\hline
\end{tabular}
\end{center}
\label{strint-tension-1}
\end{table}

Moreover, on the $24^4$ lattice~at $\beta=2.5$  
the restricted (``Abelian'') part $\sigma_{\rm rest}$ reproduces 100$\%$ of the full string tension $\sigma_{\rm full}$:
\begin{equation} 
 \frac{\sigma_{\rm rest}}{\sigma_{\rm full}} = (102 \pm 2)\% 
\quad (\text{on} \ 24^4 \ \text{lattice~at} \ \beta=2.5), 
\end{equation}
and the monopole part $\sigma_{\rm mono}$ reproduces 83$\%$ of $\sigma_{\rm rest}$: 
\begin{equation} 
 \frac{\sigma_{\rm mono}}{\sigma_{\rm rest}} = (83 \pm 1)\%  
 \Longrightarrow 
 \frac{\sigma_{\rm mono}}{\sigma_{\rm full}} = (85 \pm 2)\% 
\quad (\text{on} \ 24^4 \ \text{lattice~at} \ \beta=2.5) . 
\end{equation}
In general, the monopole part does not include  the Coulomb term and hence the linear potential 
is obtained to an accuracy better than the full potential. 
Thus, we have confirmed the {{{restricted field  dominance} (or ``Abelian'' dominance) and the {{{magnetic monopole dominance} in the string tension for the $SU(2)$ Yang-Mills theory in our framework.

\subsection{Gauge-invariant chromoelectric field and flux tube formation}

According to the dual superconductor picture for quark confinement, the QCD vacuum must be a dual superconductor so that the chromoelectric field generated by the $q\bar q$ pair is squeezed into the flux tube forming the string structure and hence the energy per unit length of the string gives the string tension of the linear potential. 
In other words, the QCD vacuum exhibits the {{{dual Meissner effect}. 
In order to confirm the dual Meissner effect, we measure the chromofield around the $q\bar q$ pair to obtain the information on the distribution or the profile of the chromoelectric field generated by the static $q\bar q$ pair.
These issues are also checked for the restricted field to examine whether or not the restricted field $V$ can reproduce the full results  obtained by the original full field $U$.

For this purpose, we must extract the chromofield in the gauge-invariant way. 
This is a nontrivial issue. 
In order to define the gauge-invariant chromofield strength tensor, we introduce the following  operator representing a gauge-invariant connected correlator   between the Wilson loop operator and a plaquette variable according to Di Giacomo, Maggiore and Olejnik \cite{Giacomo:1990,Cea:1995}:
\begin{align} 
 \rho_{_{UP}} 
:=\frac{\langle {\rm tr}(WLU_PL^{\dagger}) \rangle}{\langle {\rm tr}(W) \rangle}
- \frac{\langle {\rm tr}(U_P){\rm tr}(W) \rangle}{{\rm tr}({\bf 1})\langle {\rm tr}(W) \rangle} ,
\label{cf1-1}
\end{align}
where $W$ is the Wilson loop operator representing a pair of quark and antiquark,  $U_P$ is the plaquette variable as the probe for measuring the chromofield strength at the position of the plaquette, and $L$ is the line connecting the plaquette  $U_P$ and the Wilson loop operator $W$, which is called the Schwinger line. See Fig.~\ref{cf-fig1}. 
Here the Schwinger line $L$ is necessary to guarantee the gauge invariance of the correlator $\rho_W$. 
We must pay attention to the orientation between $U_P$ and $W$.
The above definition works for $SU(N)$ gauge group for any $N$ using ${\rm tr}({\bf 1})=N$, and we set ${\rm tr}({\bf 1})=2$ for the gauge group $SU(2)$.

\begin{figure}[ptb]
\begin{center}
\includegraphics[
height=4cm,
]
{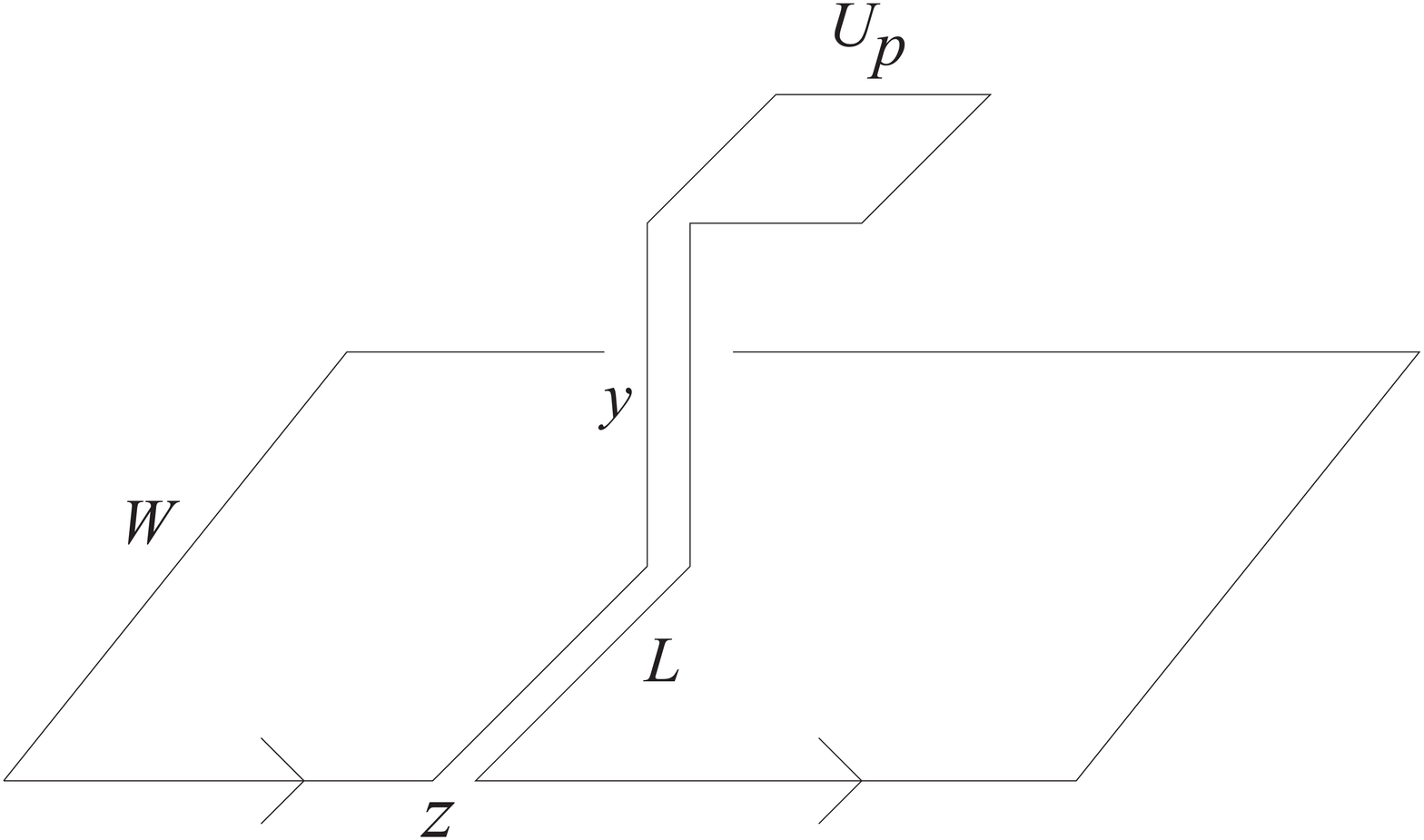} 
\end{center}
\vspace{-5mm}
\caption{
The setup of $WLU_PL^{\dagger}$ in the definition of the operator $\rho_{_{UP}}$: $z$ is the position of the Schwinger line $L$ along the line connecting $\bar{q}$ and $q$ at a fixed Euclidean time $t$, and $y$ is the distance from the plane  spanned by  the Wilson loop $W$ to the plaquette $U_P$. 
}%
\label{cf-fig1}
\end{figure}

For
$U_{x,\mu}=\exp (-ig\epsilon \mathscr{A}_\mu(x))$,  the plaquette variable is rewritten as
\begin{equation}
 U_{P}= \exp(-ig\epsilon^2 \mathscr{F}_{\mu\nu} ) 
 = {\bf 1} -ig\epsilon^2 \mathscr{F}_{\mu\nu} + O(\epsilon^4) .
\end{equation}
This leads to the trace: using the cyclicity of the trace and the unitarity $LL^\dagger=L^\dagger L={\bf 1}$, 
\begin{align}
 {\rm tr}(U_PL^{\dagger}WL) =&  {\rm tr}(L^{\dagger}WL)  -ig \epsilon^2 {\rm tr}(\mathscr{F}_{\mu\nu}L^{\dagger}WL) + O(\epsilon^4)
 \nonumber\\
=& {\rm tr}(W)  -ig \epsilon^2 {\rm tr}(\mathscr{F}_{\mu\nu}L^{\dagger}WL) + O(\epsilon^4)  ,
\end{align}
while using the traceless property   ${\rm tr}(\mathscr{F}_{\mu\nu})=0$,
\begin{equation}
 {\rm tr}(U_{P})=  {\rm tr}({\bf 1})   + O(\epsilon^4) .
\end{equation}
Hence the correlator reads
\begin{align} 
 \rho_{_{UP}} 
=  \frac{\left\langle {\rm tr}(W) \right\rangle  -ig \epsilon^2 \left\langle {\rm tr}(\mathscr{F}_{\mu\nu}L^{\dagger}WL) \right\rangle }{\langle {\rm tr}(W) \rangle} 
- \frac{ {\rm tr}({\bf 1}) \langle {\rm tr}(W) \rangle }{{\rm tr}({\bf 1})\langle {\rm tr}(W) \rangle} + O(\epsilon^4) .
\label{cf1-1b}
\end{align}
In the naive continuum limit (lattice spacing $\epsilon \to 0$), therefore, the operator $\rho_U$ reduces to the field strength in the presence of the $q\bar{q}$ source: 
\begin{equation}
\rho_{_{UP}}%
\overset{\varepsilon\rightarrow0}{\simeq}g\epsilon^{2}\left\langle
\mathscr{F}_{\mu\nu}\right\rangle _{q\bar{q}}:= - ig\epsilon^{2} \frac{\left\langle
\mathrm{tr}\left(   \mathscr{F}_{\mu\nu}L^{\dag}WL \right)
\right\rangle }{\left\langle \mathrm{tr}\left(  W\right)  \right\rangle
}+O(\epsilon^{4}) .
\label{cf1-2}
\end{equation}
Therefore, we can define a gauge-invariant chromofield strength tensor by
\begin{align} 
F_{\mu\nu}[U](x)  :=  \epsilon^{-2} \frac{\sqrt{\beta}}{2} \rho_U(x), \quad \beta := \frac{2N}{g^2} \quad (\text{for} \ G=SU(N)) . 
\label{cf1-3}
\end{align}

In the definition of the operator $\rho_{_{UP}}$,  $WLU_PL^{\dagger}$ is set up as follows.
Let $z$ be the position of the Schwinger line $L$ along the line connecting $\bar{q}$ and $q$ at a fixed Euclidean time $t$, and $y$ be the distance from the plane spanned by the Wilson loop $W$ to the plaquette $P$. See Fig.~\ref{cf-fig1}.
By changing the distances $y,z$ and the direction of the plaquette $U_P$ relative to the Wilson loop $W$, 
we can scan the chromoelectric and chromomagnetic fields around the $q\bar{q}$ pair.

Similarly, we define the chromofield strength tensor $F_{\mu\nu}[V]$ from  the restricted field $V_{\mu}(x)$ by
\begin{align} 
F_{\mu\nu}[V](x)  &:=  \epsilon^{-2} \frac{\sqrt{\beta}}{2} \rho_{_{V}}(x), \quad 
\rho_{_{VP}}  :=  \frac{\langle {\rm tr}(W^VL^VV_P{L^V}^{\dagger}) \rangle}{\langle {\rm tr}(W^V) \rangle}
-\frac{1}{{\rm tr}({\bf 1})}\frac{\langle {\rm tr}(V_P){\rm tr}(W^V) \rangle}{\langle {\rm tr}(W^V) \rangle} ,
\label{cf1-5}
\end{align}
where $V_P$ is the plaquette variable for the restricted field (link variable) $V$, $W^V$ and $L^V$ represent respectively the Wilson loop operator and the Schwinger line constructed from the restricted field (link variable) $V$.



In the numerical simulations,  we have generated the link fields ${U_{x,\mu}}$ using the heat bath method for the standard $SU(2)$ Wilson action.
We have stored 100 configurations for the  
  $24^4$ lattice at $\beta=2.5$ with 100 iterations. 
We take $R=T=8$ for the size of the Wilson loop operator to calculate the operators (\ref{cf1-3}) and (\ref{cf1-5}). 
Therefore, the quark and antiquark source is introduced as $R \times T$ Wilson loop $W$ in the $z$-$t$ plane.
The probe $U_{P}$ is set at the center of the Wilson loop and moved along the $y$-direction. 
We have performed the {{{hypercubic blocking (HYP)} \cite{Hasenfratz:2001} as a smearing method to obtain  ${U_{x,\mu}}$ for calculating the operators (\ref{cf1-3}) and (\ref{cf1-5}).
See the Appendix for the details of the HYP.


\begin{figure}[ptb]
\begin{center}
\includegraphics[scale=0.5]{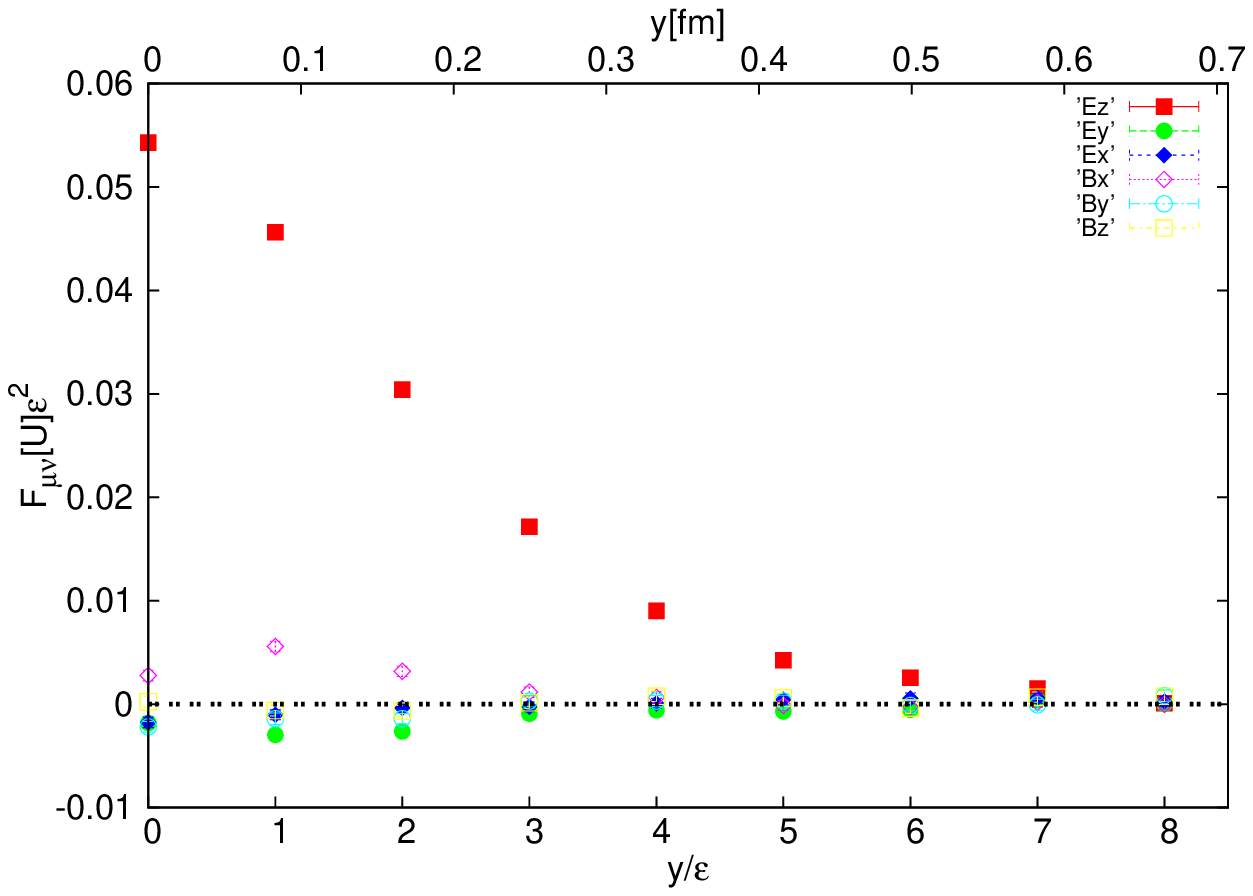} 
\quad
\includegraphics[scale=0.7]{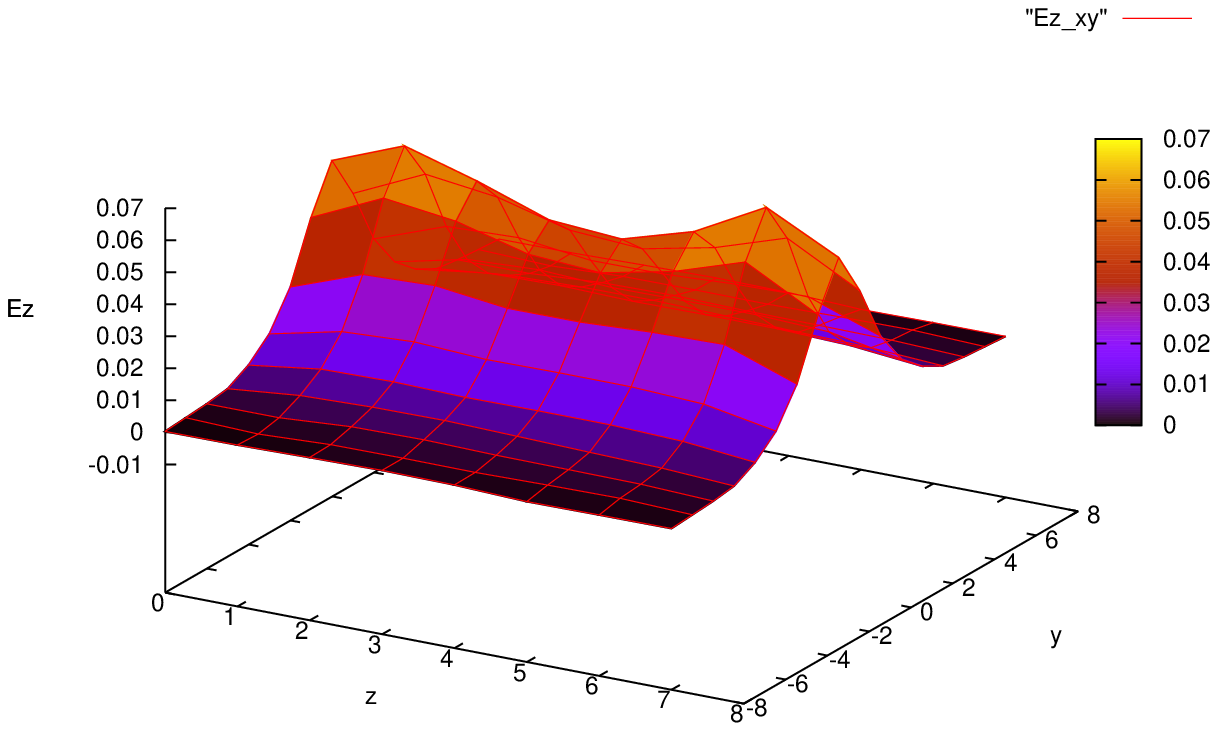}
\vspace{-5mm}
\end{center}
\caption{
The chromoelectric and chromomagnetic fields obtained from the full field $U$ on $24^4$ lattice at $\beta=2.5$.
(Left panel) 
$y$ dependence of the chromoelectric field $E_i(y)=F_{4i}(y)$ ($i=x,y,z$) at fixed $z=4$ (mid-point of $q\bar q$). 
(Right panel)
The distribution of $E_z(y,z)$ obtained for the $8 \times 8$ Wilson loop with $\bar{q}$ at $(y,z)=(0,0)$ and $q$ at $(y,z)=(0,8)$.
}
\label{cf-fig2}
\end{figure}

The results of numerical simulations are shown in Fig.~\ref{cf-fig2}.
In the left panel of Fig.~\ref{cf-fig2}, we find that only the $E_{z}$ component of the chromoelectric field $(E_x,E_y,E_z)=(F_{14},F_{24},F_{34})$ connecting $q$ and $\bar q$ has non-zero value for the original Yang-Mills field $U$.
The other components are zero consistently within the numerical errors. 
In other words, the chromoelectric field is directed to the line connecting  quark and antiquark.
The magnitude of the chromoelectric field $E_{z}$   decreases quickly as the distance $y$ increases in the direction perpendicular to the line.
Thus the obtained profile of the chromoelectric field represents the structure expected for the flux tube.
Therefore, we have confirmed the formation of the chromoelectric flux in $SU(2)$ Yang-Mills theory on a lattice.

To see the profile of the non-vanishing component $E_z$ of the chromoelectric field in detail, we explore the distribution of chromoelectric field on the 2-dimensional plane. 
The right panel of Fig.~\ref{cf-fig2} shows the distribution of $E_{z}$ component of the chromoelectric field, where the quark-antiquark source represented as the $R \times T$ Wilson loop $W$ is placed at $(Y,Z)=(0,R), (0,0)$, and the probe $U_P$ is displaced on the $Y$-$Z$ plane at the midpoint of the $T$-direction. 
The magnitude of $E_{z}$ is shown by the height of the 3D plot.
We find that the magnitude $E_{z}$  is almost uniform for the original part $U$ except for the neighborhoods of the locations of $q$, $\bar q$ source.

Next, the results for the restricted field $V$ is shown in Fig.~\ref{cf-fig3}.  
From the left panel of Fig.~\ref{cf-fig3}, we find that the strength of the chromoelectric field obtained from the  restricted field becomes smaller than the full one, but the structure of the flux tube survives. 
The ratio of the flux at the origin   $y=0$ is  
\begin{align}
 E^U_z(0)=5.289 \times 10^{-2} , \ E^V_z(0)=3.965\times 10^{-2}, \ E^V_x(0)/E^U_x(0)=0.723 
\quad (\text{on} \ 24^4 \ \text{lattice~at} \ \beta=2.5), 
\end{align}
From the right panel of Fig.~\ref{cf-fig3}, we find that the magnitude $E_{z}$ is quite uniform for the restricted field $V$, compared with the full field. 
This difference is due to  the contributions from the remaining part $X$ which affects only the  short distance, as the correlator of the $X$ field exhibits the exponential fall-off and disappears quickly in the distance as shown in \cite{SKKMSI07}. 
 Thus the restricted field $V$ reproduces the chromoelectric flux tube in the $SU(2)$ Yang-Mills theory on a lattice.


\begin{figure}[ptb]
\begin{center}
\includegraphics[scale=0.5]{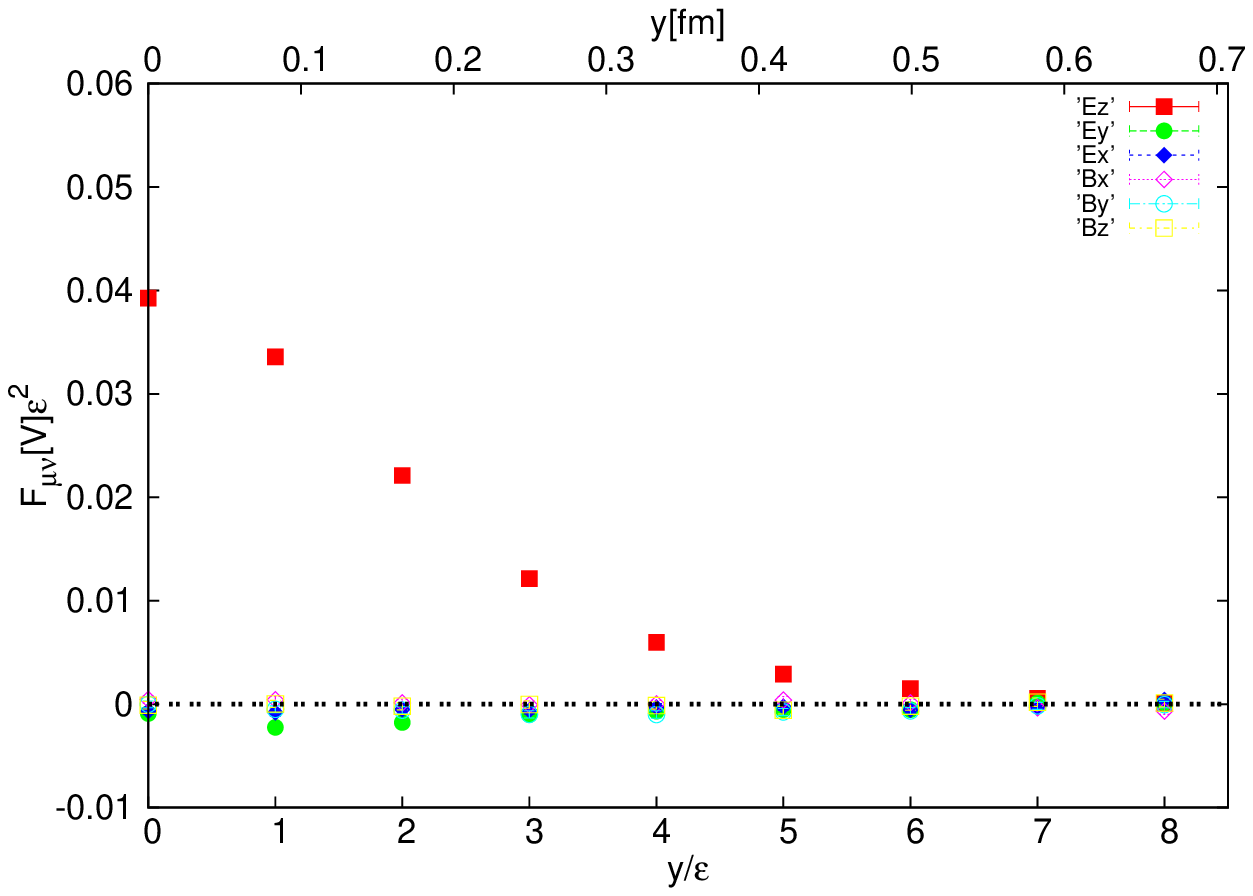} 
\quad
\includegraphics[scale=0.7]{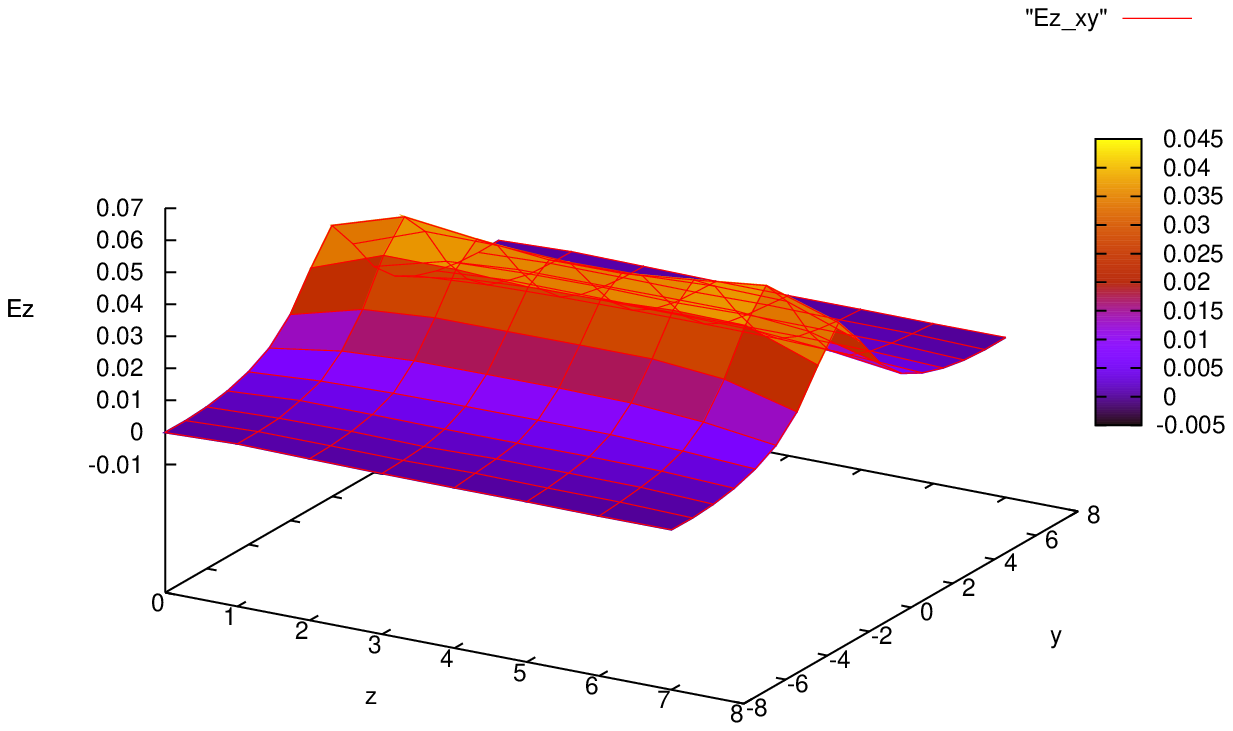} 
\vspace{-5mm}
\end{center}
\caption{
The chromoelectric field obtained from the restricted field $V$ on  $24^4$ lattice at $\beta=2.5$. 
}
\label{cf-fig3}
\end{figure}

For comparison, we have calculated also the operator which was estimated by Di~Giacomo et al.\cite{Giacomo:1990}:
\begin{align} 
 \rho_U' &= \frac{\langle {\rm tr}(U_PL^{\dagger}WL) \rangle}{\langle {\rm tr}(W) \rangle} - \frac{\langle {\rm tr}(U_P) \rangle}{{\rm tr}({\bf 1})}, 
\quad
F'_{\mu\nu}(x)  = \frac{\sqrt{\beta}}{2} \rho_U'(x) 
 .
\label{cf2-1}
\end{align}
It is easy to see that the operator $\rho'$ has the same expression as (\ref{cf1-2}) up to the order $O(\epsilon^2)$ and the difference appears in the order $O(\epsilon^4)$.
The result is shown in Fig.~\ref{cf-fig2-2}.  The comparison of Fig.~\ref{cf-fig2-2} with the left panel of Fig.~\ref{cf-fig2} shows that the value of (\ref{cf2-1}) is consistent with (\ref{cf1-1}).  
The numerical data are given in Table~\ref{cf-tbl2-1}.

\begin{figure}[ptb]
\begin{center}
\includegraphics[scale=0.5]{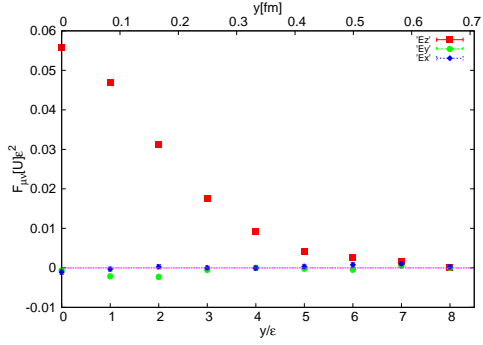} 
\end{center}
\vspace{-5mm}
\caption{
The chromoelectric field $E_j(y)=F_{4j}(y)$ ($j=x,y,z$) obtained from  (\ref{cf2-1}) as a function of the distance $y$, 
on $24^4$ lattice at $\beta=2.5$. 
}%
\label{cf-fig2-2}
\end{figure}


\begin{table}[hptb]
\caption{
The comparison of the chromoelectric field obtained from 
(\ref{cf1-1}) and (\ref{cf2-1}) on $24^4$ lattice at $\beta=2.4$.
We fix $z$ to be the midpoint, i.e., $z=4$.
}
\label{cf-tbl2-1}
\begin{center}
\begin{tabular}{ccc}\hline
y & $E_z(y)=F_{41}(x)$ &  $E_z'(y)=F'_{41}(x)$ \\ \hline
0 &  $5.428(\pm0.062)\times10^{-2} $ & $5.585(\pm0.065)\times10^{-2}$ \\
1 &  $4.560(\pm0.055)\times10^{-2}$ & $4.699(\pm0.059)\times10^{-2}$ \\
2 &  $3.041(\pm0.055)\times10^{-2}$ & $3.127(\pm0.058)\times10^{-2}$ \\
3 &  $1.714(\pm0.050)\times10^{-2}$ & $1.751(\pm0.053)\times10^{-2}$ \\
4 &  $0.901(\pm0.049)\times10^{-2}$ & $0.914(\pm0.054)\times10^{-2}$ \\
5 &  $0.424(\pm0.047)\times10^{-2}$ & $0.426(\pm0.051)\times10^{-2}$ \\
6 &  $0.255(\pm0.048)\times10^{-2}$ & $0.251(\pm0.051)\times10^{-2}$ \\
7 &  $0.149(\pm0.050)\times10^{-2}$ & $0.157(\pm0.054)\times10^{-2}$ \\
8 &  $0.009(\pm0.049)\times10^{-2}$ & $0.011(\pm0.052)\times10^{-2}$ \\\hline
\end{tabular}
\end{center} 
\end{table}

\subsection{Magnetic current and  dual Meissner effect}

Although we have confirmed the formation of the chromoelectric flux in $SU(2)$ Yang-Mills theory on a lattice, the existence of the flux tube alone is not sufficient for proving the occurrence of the dual Meissner effect. 

Next, we investigate the relation between the chromoelectric flux and the  magnetic current. 
The magnetic(-monopole) current can be calculated as
\begin{equation}
 k= \delta{}^{\displaystyle *}F[V] ={}^{\displaystyle *}d F[V] , 
\label{def-k}
\end{equation}
where $F[V]$ is the field strength (\ref{cf1-5}) defined from the the restricted field $V$  in the presence of the $q\bar q$ source,
$d$
the exterior derivative, $\delta$ codiffrential, and $^{\ast}$ denotes the Hodge dual operation. 
Note that non-zero magnetic current follows from violation of the Bianchi identity  
(If the field strength was given by the exterior derivative of some  field $A$ (one-form), $ F=dA$, \ we would obtain $k=\delta{}^{\displaystyle *}F={}^{\displaystyle *}d^{2}A=0$).

If only the components $E_z=F_{34}=-F_{43}$ are non-vanishing among $F_{\alpha\beta}$, then $k^{\mu}   =  \frac{1}{2}\epsilon^{\mu\nu\alpha\beta} \partial_{\nu}F_{\alpha\beta}$ reads 
\begin{align}
k^{\mu}  = \epsilon^{\mu\nu34} \partial_{\nu}F_{34}
= \epsilon^{\mu\nu34} \partial_{\nu}E_{z}   ,
\end{align}
and the non-vanishing components of $k^{\mu}$ are  given by  $k^1, k^2$ in the $X$-$Y$ plane:
\begin{align}
k^{1} 
= \partial_{y}E_{z}, \quad
k^{2} 
= - \partial_{x}E_{z},  \quad k^3=0,  \quad k^4=0 .
\end{align}

\begin{figure}[ptb]
\begin{center}
\includegraphics[scale=0.55]{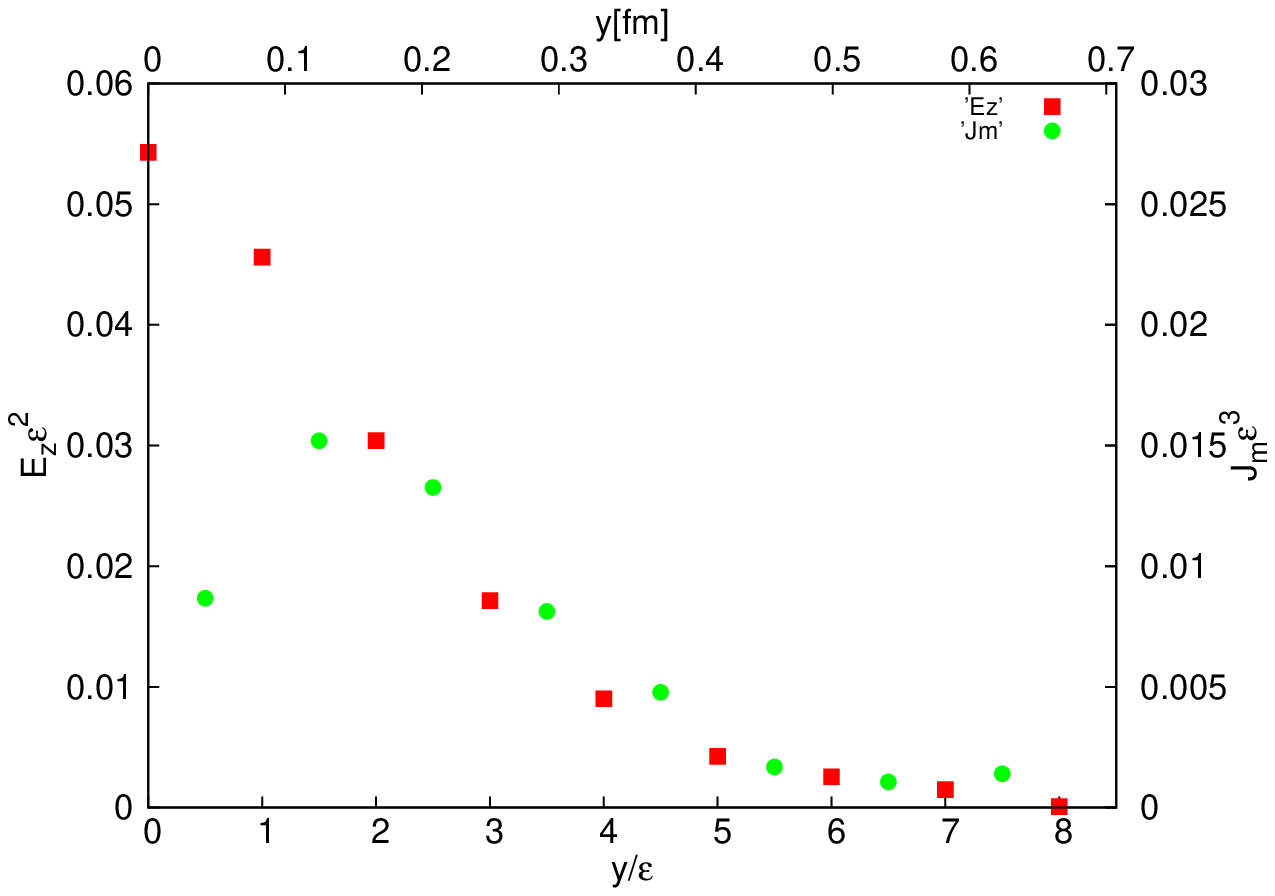}
\includegraphics[scale=0.65]{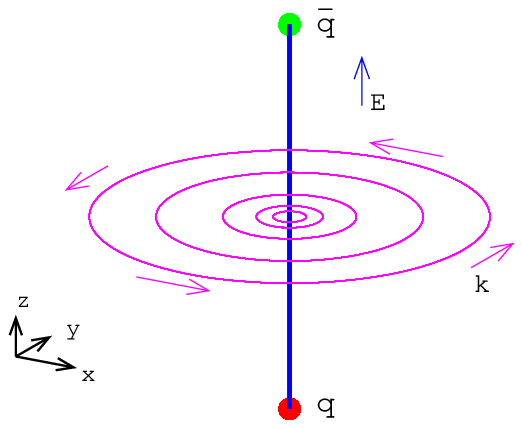} 
\includegraphics[scale=0.55]{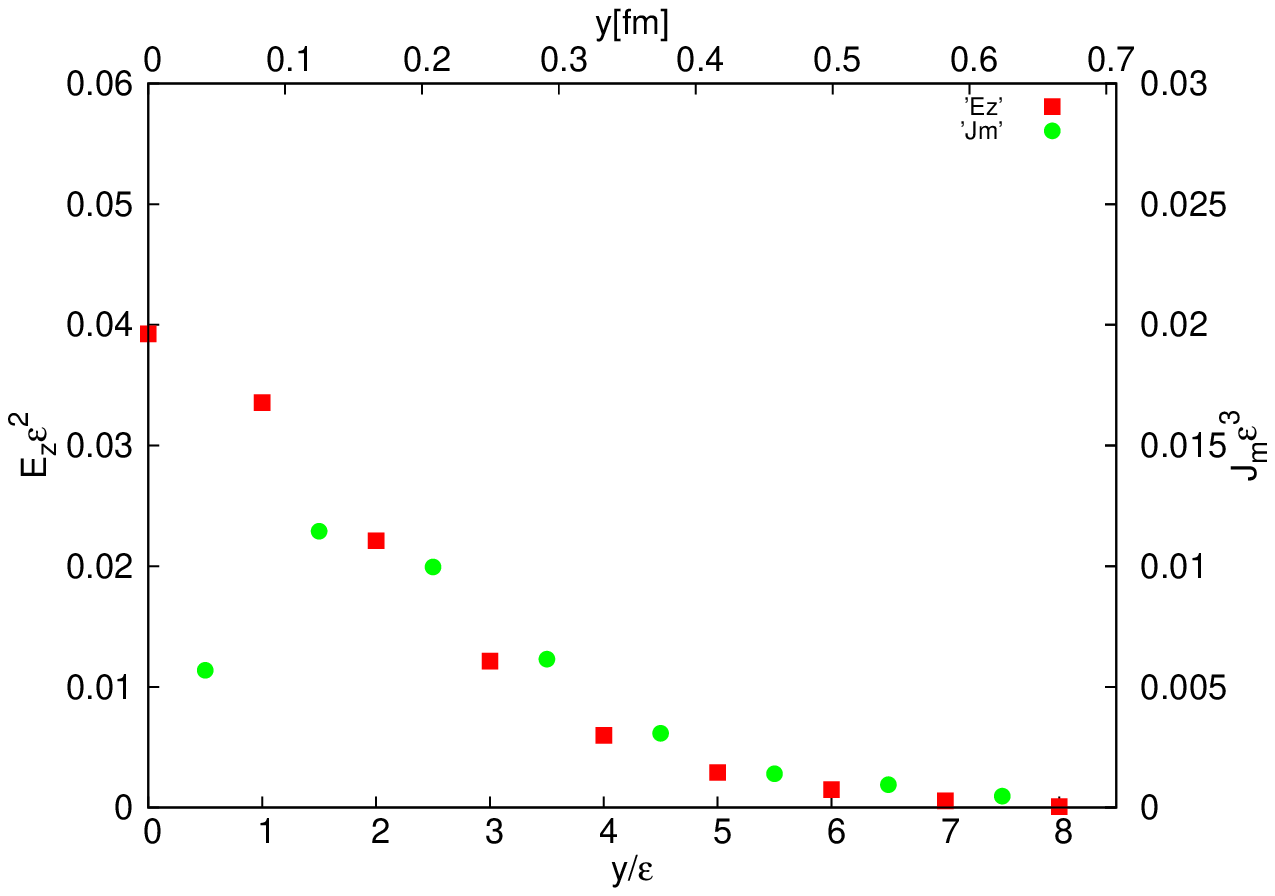}
\vspace{-5mm}
\end{center}
\caption{{}
The magnetic-monopole current $\mathbf{k}$ induced around the chromoelectric flux along the $z$ axis connecting a  pair of quark and antiquark.
(Center panel) 
The positional relationship between the chromoelectric field $E_{z}$ and the magnetic current $\mathbf{k}$. 
(Left panel) 
The magnitude of the chromoelectric field $E_{z}$ and the magnetic current  $J_{m}=|\mathbf{k}|$ as functions of the distance $y$ from the $z$ axis calculated from the original full variables. 
(Right panel) 
The counterparts of the left graph calculated from the restricted variables. 
}
\label{fig:Mcurrent}%
\end{figure}

Fig.~\ref{fig:Mcurrent} shows the  magnetic current measured in $X$-$Y$ plane at the midpoint of $q\bar q$ pair in the $Z$-direction. 
The left panel of Fig.~\ref{fig:Mcurrent} shows the positional relationship between chromoelectric flux and  magnetic current.
The right panel of Fig.~\ref{fig:Mcurrent} shows the magnitude of the  chromoelectric field $E_z$ (left
scale) and the magnetic current $k$ (right scale). 
The existence of non-vanishing magnetic current $k$ around the chromoelectric field $E_z$ supports the dual picture of the ordinary superconductor exhibiting the electric current $J$ around the magnetic field $B$.

The above results show the simultaneous formation of the chromoelectric flux tube and the associated magnetic-monopole current induced induced around it.    
Thus, we have confirmed the dual Meissner effect in $SU(2)$ Yang-Mills theory on a lattice. 
We have also shown that the restricted field $V$ reproduces the dual Meissner effect in the $SU(2)$ Yang-Mills theory on a lattice.

In our formulation, it is possible to define a gauge-invariant magnetic-monopole current  $K_{\mu}$  by using $V$-field,
\begin{subequations}
\begin{align}
K_{\mu}(x)  &  =2\pi m_{\mu}(x) :=\frac{1}{2}\epsilon_{\mu\nu\alpha\beta} \partial_{\nu} \bar\Theta_{\alpha\beta}(x) ,
 \\
\bar\Theta_{\mu\nu}(x)   &  :=-\arg \left\{ \text{ \textrm{tr}}\left[  \left(   \mathbf{1}+\bm{n}_{x}\right)  V_{x,\mu}V_{x+\mu,\mu}V_{x+\nu,\mu}^{\dag}V_{x,\nu}^{\dag} \right]/{\rm tr}(\bm{1}) \right\} ,
\end{align}
\end{subequations}
which is obtained from the field strength $\mathscr{F}[\mathscr{V}]$ of the restricted field $\mathscr{V}$, as suggested from the non-Abelian Stokes theorem \cite{Kondo08}.
The magnetic-monopole current  $K_{\mu}$ defined in this way can be used to study the magnetic current around the chromoelectric flux tube, instead of the above definition   (\ref{def-k}) of $k$.
The comparison of two monopole currents $k$ will be done in the forthcoming paper.

\subsection{Ginzburg-Landau parameter and type of dual superconductor}

Moreover, we investigate the type of the dual superconductor in the QCD vacuum. 
The {{{Ginzburg-Landau (GL) parameter} in the superconductor is defined from the penetration depth $\lambda$ and the coherence length $\xi$ by
\begin{align} 
\kappa = \frac{\lambda}{\xi} .
\label{cf4-1}
\end{align}
The superconductor is called the type-I when $\kappa<\frac{1}{\sqrt{2}}$, while type-II when $\kappa>\frac{1}{\sqrt{2}}$.
In the type-I superconductor, the attractive force acts between two vortices, while the repulsive force in the type-II superconductor.
There is no interaction at $\kappa=\frac{1}{\sqrt{2}} \simeq 0.707$. 
The preceding studies support that the dual superconductor for the $SU(2)$ lattice Yang-Mills theory is at the border between type-I and type-II, or  weak type-I \cite{Suzuki:1988}.

Usually, in the dual superconductor of the type II, it is justified to use the asymptotic form $K_0(y/\lambda)$ to fit the chromoelectric field in the large $y$ region (as the solution of the Ampere equation in the dual GL theory).  
However, it is clear that this solution cannot be applied to the small $y$ region, as is easily seen from the fact that $K_0(y/\lambda) \to \infty$ as $y \to 0$. 
In order to see the difference between type I and type II, it is crucial to see the relatively small $y$ region.
Therefore, such a simple form cannot be used to detect the type I dual  superconductor. 
However, this important aspect was ignored in the preceding studies except for a work \cite{CCP12}.

We proceed to determine the GL parameter $\kappa$ of the dual superconductor for $SU(2)$ lattice Yang-Mills theory using the numerical data for the chromoelectric field obtained in the previous section.
We can measure the penetration depth $\lambda$ of the chromoelectric field directly from the data obtained in the previous section without any assumption. 
In order to obtain the the coherence length $\xi$, however, we must solve the coupled nonlinear differential equations in the {{{GL theory}, i.e., the {{{GL equation} and the  Ampere equation. 
In the GL theory, the gauge field $A$ and the scalar field $\phi$ obey simultaneously  the GL equation:
\begin{equation}
 (\partial^\mu -iq A^\mu)(\partial_\mu -iq A_\mu) \phi + \lambda_4 (\phi^* \phi - \eta^2) = 0 ,
\end{equation}
and the Ampere equation:
\begin{equation}
 \partial^\nu F_{\mu\nu} + iq [\phi^* (\partial_\mu \phi -iq A_\mu \phi)  - (\partial_\mu \phi -iq A_\mu \phi)^* \phi] = 0 .
\end{equation}
To avoid this, we follow the method given by Clem \cite{Clem75} invented for the ordinary superconductor based on the GL theory, which was recently applied to the dual superconductor for $SU(3)$ lattice Yang-Mills theory by  
\cite{CCP12,SKKS12}.
The advantage of this method is that it is able to take into account the whole range of $y$ for fitting the data to determine precisely the type of (dual) superconductivity, in sharp contrast to the preceding approach which uses only the asymptotic region at large $y \gg 1$. 
By applying the Clem method to the dual superconductor, the chromoelectric field $E_z(y)$ must obey
\begin{align} 
E_z(y) = \frac{\Phi}{2\pi}\frac{\mu^2}{\alpha}\frac{K_0(\sqrt{\mu^2y^2+\alpha^2})}{K_1(\alpha)} ,
\label{cf4-2}
\end{align}
where $\Phi$ is the external electric flux, $\mu$ and $\alpha$ are defined by 
\begin{align} 
\mu := \frac{1}{\lambda}, \quad {\alpha} := \frac{\zeta}{\lambda} ,
\label{cf4-3}
\end{align}
and $K_0$ and $K_1$ are the modified Bessel functions of zeroth and first order respectively. 
Here $\zeta$ is the variational parameter representing  the core radius. 
The GL parameter $\kappa$ is written in terms of $\alpha$ alone:
\begin{align} 
\kappa = \frac{\sqrt{2}}{\alpha}[1-K_0^2(\alpha)/K_1^2(\alpha)]^{1/2}, \quad {\alpha} := \frac{\zeta}{\lambda} .
\label{cf4-4}
\end{align}

Then the obtained value of $\alpha$ is used to determine the GL parameter $\kappa$ according to (\ref{cf4-4}).

 


The graph of the fitting is given in Fig.~\ref{cf-fig5} and the obtained values for the fitted parameters are given in Table~\ref{cf-tbl4-1}
where we have used the fitting function: 
\begin{equation}
E_z(y)=c K_{0}(\sqrt{\mu^{2}y^{2}+\alpha^{2}}) , \quad
c= \frac{\Phi}{2\pi}\frac{\mu^2}{\alpha}\frac{1} {K_{1}(\alpha)}
= \frac{\Phi}{2\pi}\frac{1}{\lambda \zeta}\frac{1} {K_{1}(\zeta/\lambda)} .
\label{fitting}
\end{equation}
Thus we have obtained the GL parameter for the full field $\kappa_U$ and the restricted field $\kappa_V$:
\begin{align} 
\kappa_U = 0.484 \pm 0.070 \pm 0.026, \quad \kappa_V = 0.377 \pm 0.079 \pm 0.018.
\label{cf4-7}
\end{align}
Here and in what follows, the first error denotes the statistics error and the second one denotes the systematic error or the lattice artifact due to choosing the center or the corner of the plaquette as the representative of $E_z(y)$.


\begin{figure}[ptb]
\begin{center}
\includegraphics[scale=0.6]{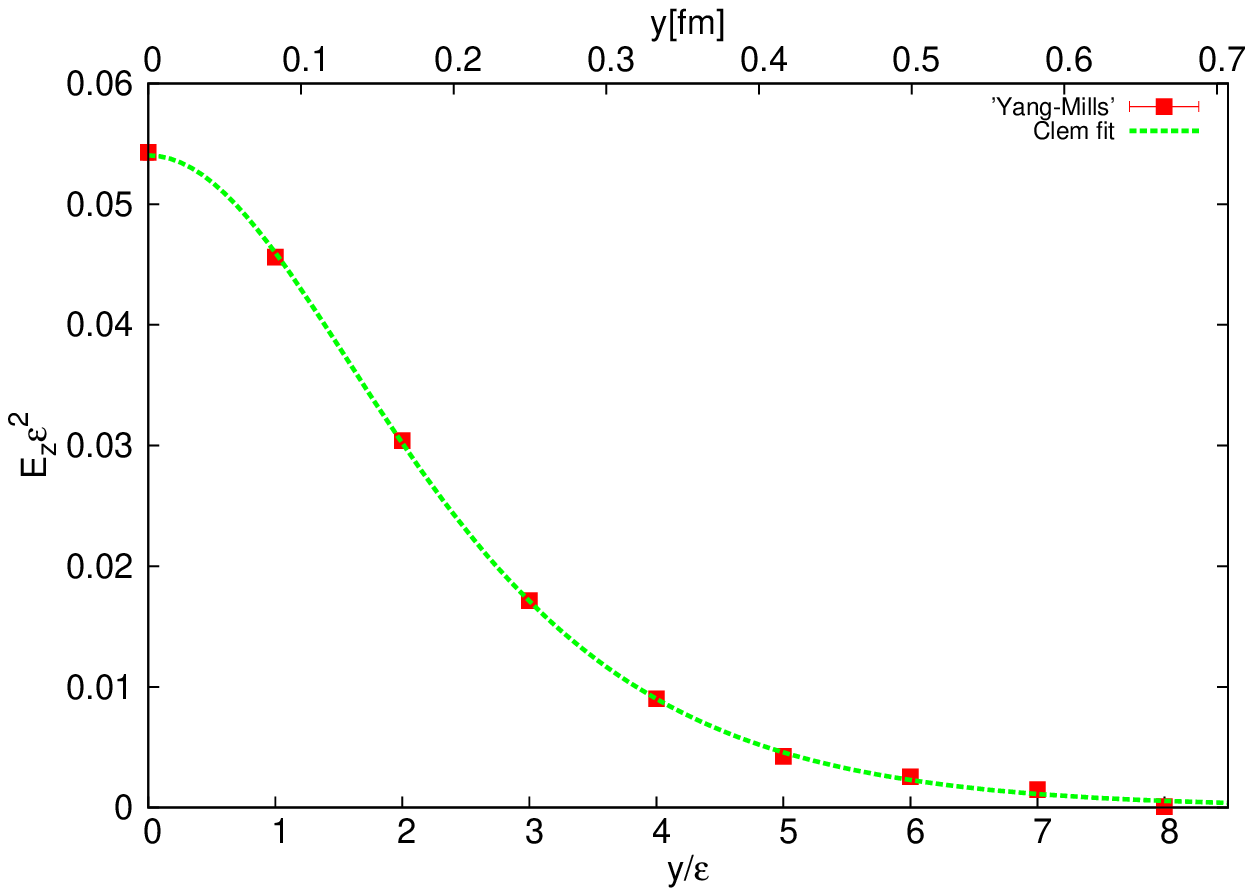} 
\quad
\includegraphics[scale=0.6]{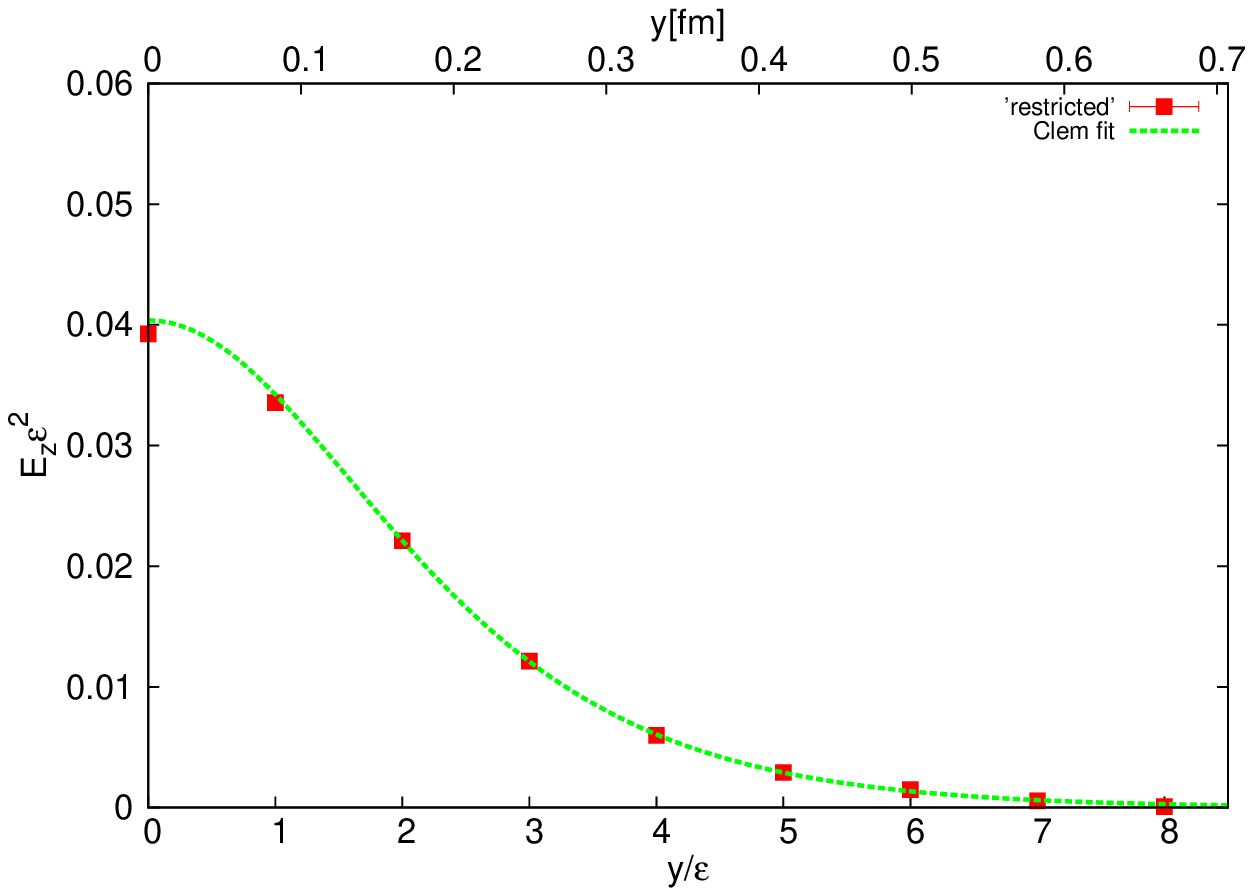} 
\vspace{-5mm}
\end{center}
\caption{
The magnitude of the chromoelectric field $E_z(y)$ as a function of the distance $y$. 
(Left panel) The data of Fig.~\ref{cf-fig2} and the fitted result for the full field, 
(Right panel) The data of Fig.~\ref{cf-fig3} and the fitted result for the restricted field.
The fit range we have used are [0,8] for the full field in the left panel and [2,8] for the restricted field in the right panel. 
}
\label{cf-fig5}
\end{figure}

\begin{table}[htp]
\caption{Summary of the fit values }
\label{cf-tbl4-1}
\begin{center}
\begin{tabular}{lccc}\hline
                                   & $c$    &  $\mu$  &  $\alpha$  \\ \hline
link field $U$           &  $0.355 (0.096)$  & $0.689 (0.039)$  &  $1.767 (0.218)$    \\
restricted field $V$ &  $0.414 (0.187)$  & $0.774 (0.052)$  &  $2.128 (0.400)$    \\ \hline
\end{tabular}
\end{center} 
\end{table}



The penetration depth $\lambda$ is obtained  using the first equation of (\ref{cf4-3}) from $\mu$, i.e., $\lambda=1/\mu$, while the coherence length $\xi$ is obtained using (\ref{cf4-1}) from $\lambda$ and the GL parameter $\kappa$ (\ref{cf4-7}), i.e., $\xi=\lambda/\kappa$. 
The full link variable yields
\begin{align} 
\lambda_U = 0.121(7+0) \text{fm}, \quad \xi_U = 0.250(4+1) \text{fm}.
\label{cf4-8a}
\end{align}
where we have used the value of scale 
 $\epsilon(\beta=2.5)=0.08320$ fm  of Ref.\cite{SKKMSI07}.
While the restricted field gives
\begin{align} 
\lambda_V = 0.107(7+0) \text{fm}, \quad \xi_V = 0.285(7+1) \text{fm} .
\label{cf4-8b}
\end{align}
The obtained results are consistent with the result $\lambda=0.1135(27)\text{fm}$ of Ref.\cite{CCP12}.

Our results show  that the dual superconductor for the $SU(2)$ lattice Yang-Mills theory is the weak type I, rather than the border between type-I and type-II.
Our results are to be compared with the preceding works.
First, Cea, Cosmai and Papa \cite{CCP12} have used the same operator and the same fitting function (\ref{cf4-2}) as ours.  In the calculation of the operator, however, they used the cooling method, while we have used HYP. 
In addition, they have confirmed that the GL parameter is shifting towards the type I as the cooling step is forward.
They used the constant fit of the data obtained at three points $\beta=2.252,2.55,2.6$ on a lattice $20^4$, and they checked that the value of $\kappa$ does not depend on $\beta$. 
The obtained value of the GL parameter $\kappa_U$ is consistent with the value $\kappa=0.467\pm 0.310$  within errors.
From our point of view, the cooling method cannot be applied to the restricted part, since it amounts to changing the action. 

Next, Bali, Schlichter and Schilling \cite{Suzuki:1988} have used the different operator without the Schwinger line, namely, the action distribution proportional to the squared field strength, and applied APE smearing to the Wilson loop alone.
They have used a different fitting function than ours.
The result is $\kappa=0.59^{+13}_{-14}$  indicating the border between type I and II.

Third, Suzuki et al. \cite{Suzuki:2009xy} have used the Abelian-projected operator and applied the APE smearing to the Wilson loop alone.
They used the improved Iwasaki action at three values of $\beta=1.10,1.28,1.40$ to calculate  the Wilson loop of small size: $W(R=3,T=5)$, $W(R=5,T=5)$, $W(R=7,T=7)$ where the interquark distance $q-\bar{q}$ is fixed to be 0.32fm.
Moreover, $\langle E_{A_z}(y) \rangle_W$ is fitted to $c_1\exp(-y/\lambda)+c_0$, and $\langle  k_{\mu}^2(y) \rangle_W$ is fitted to $c'_1\exp(-\sqrt{2}y/\xi)+c'_0$ to calculate the GL parameter. Their result is $\sqrt{2} \kappa=1.04(7), 1.19(5), 1.09(8)$ indicating the border between type I and II.

\subsection{A new gauge-invariant chromofield strength}

An advantage of the new formulation is that we can give another definition of the gauge-invariant chromofield strength in the presence of the $q\bar q$ source, which does not need the Schwinger line $L$ and $L^\dagger$ to give the gauge-invariant chromofield strength.
We propose a gauge-invariant chromofield strength:
\begin{align} 
 \tilde\rho_V  &= \frac{\langle \epsilon^2 \bar{\Theta}_{x,\mu\nu}[V, \bm{n}] {\rm tr}(W) \rangle}{\langle {\rm tr}(W) \rangle}  
 \overset{\varepsilon\rightarrow0}{\simeq}g\epsilon^{2}\left\langle
\mathscr{F}_{\mu\nu}\right\rangle _{q\bar{q}}:= \frac{\langle \epsilon^2 \bar{\Theta}_{x,\mu\nu}[V,\bm{n}] {\rm tr}(W) \rangle}{\langle {\rm tr}(W) \rangle}  +O(\epsilon^{4}) ,
\quad
\tilde F_{\mu\nu}^{V}(x)  = \frac{\sqrt{\beta}}{2} \tilde\rho(x) 
  ,
\label{F-new1}
\end{align}
where 
\begin{align} 
\bar{\Theta}_{x,\mu\nu}[V,\bm{n}] := \epsilon^{-2}
{\rm arg} ( {\rm tr} \{({\bf 1}+ \bm{n}_x)V_{P} \}/{\rm tr}({\bf 1})) .
\label{cfn-mono-5b}
\end{align}
Since $\bar{\Theta}_{x,\mu\nu}[V,{\bf n}]$ is   gauge-invariant from the beginning, we do not need the Schwinger line $L$ and $L^\dagger$ to define gauge-invariant chromofield strength.
Note that $\tilde\rho_V$ is equal to 
\begin{align} 
 \tilde\rho_V  &= \frac{\langle {\rm tr} \{({\bf 1}+\bm{n}_x)V_{P} \}  {\rm tr}(W) \rangle}{\langle {\rm tr}(W) \rangle}  
 .
\label{F[V]-new2}
\end{align}
In view of this, we can also define the gauge-invariant field strength related to the original variable:
\begin{align} 
 \tilde\rho_U  &= \frac{\langle {\rm tr} \{({\bf 1}+ \bm{n}_x)U_{P} \}  {\rm tr}(W) \rangle}{\langle {\rm tr}(W) \rangle}  
\overset{\varepsilon\rightarrow0}{\simeq}g\epsilon^{2}\left\langle
\mathscr{F}_{\mu\nu}[U]\right\rangle _{q\bar{q}}:= \frac{\langle \epsilon^2 \bar{\Theta}_{x,\mu\nu}[U,\bm{n}] {\rm tr}(W) \rangle}{\langle {\rm tr}(W) \rangle}  +O(\epsilon^{4}) ,
\quad
\tilde F_{\mu\nu}^{U}(x)  = \frac{\sqrt{\beta}}{2} \tilde\rho_U(x) 
 .
\label{F[U]-new3}
\end{align}
Although the numerical simulations based on these  operators are in principle possible, the detailed studies  will be postponed to the subsequent works.

\section{Conclusion}

In this paper, we have shown numerically a gauge-independent restricted-field (``Abelian'')  dominance and magnetic-monopole dominance in the string tension  extracted from the Wilson loop average.
In particular, the data presented in this paper demonstrate that the restricted field (``Abelian'') dominance becomes complete, i.e., 100\% within the errors, while the monopole dominance is less dominant and is expected to be improved in the continuum limit. 
This result has been obtained in the gauge-independent way  based on a new formulation of  the Yang-Mills theory on a lattice, which reduces to the new variables of Cho-Duan-Ge-Faddeev-Niemi in the continuum limit. 
It should be remarked that the Abelian dominance and magnetic-monopole dominance  have been so far shown only in a special Abelian gauge fixing called  MA gauge which breaks the color symmetry explicitly. 

Moreover, we have investigated the dual Meissner effect and the type of the dual superconductor which is characterized by the Ginzburg-Landau parameter according to the Ginzburg-Landau theory. 
Our result shows that the dual superconductor for the $SU(2)$ lattice Yang-Mills theory is the border between type-I and type-II, which is consistent with the preceding results \cite{Suzuki:1988,CCP12}, or rather the weakly type I. 
We have confirmed that the same conclusion can be reproduced by the restricted field  on the type of dual superconductor for the $SU(2)$ lattice Yang-Mills theory.
These results establish the existence of the dual Meissner effect and the resulting dual superconductor mechanism  in the $SU(2)$ lattice Yang-Mills theory  in the gauge-independent way, which is responsible for quark confinement.

It should be remarked that the flux tube formation alone is not sufficient for proving the occurrence of the dual Meissner effect. 
In our works, therefore, the dual Meissner effect is examined by the simultaneous formation of the chromoelectric flux tube and the associated magnetic-monopole current induced around it.    
In a quite recent work \cite{Shibata-lattice14}, moreover, we have shown that  at finite temperature the chromoelectric flux becomes broader and other components of the flux appear and that the associated magnetic-monopole current vanishes at the critical temperature. 
This indicates the dual Meissner effect disappears above the critical temperature, which is detected by a set of flux tube and the associated magnetic-monopole current.   
The magnetic monopole current is more sensitive to the appearance or disappearance of the dual superconductivity.

In oder to draw the definite conclusion on physical quantities in the continuum limit, e.g., the Ginzburg-Landau parameter, however, we must study the scaling of the data obtained in the numerical simulations. For this purpose, we need to accumulate more data at various choices for the gauge coupling on the lattices with different sizes. 
These results will be given in a forthcoming paper. 
In the future, moreover, we hope to study the electric-current contribution to the Wilson loop average and the Abelian dominance and monopole dominance in the adjoint Wilson loop with the  possibilities of their connections to the Casimir scaling and string breaking.

\subsection*{Acknowledgement}

This work is financially supported in part by Grant-in-Aid for Scientific Research (C) 24540252 from Japan Society for the Promotion of Science (JSPS).
This work is in part supported by the Large Scale Simulation Program 
No.09-15 (FY2009), No.T11-15 (FY2011), No.12/13-20 (FY2012-2013) and No.13/14-23 (FY2013-2014) of High Energy Accelerator Research Organization (KEK).

\appendix
\section{Hypercubic blocking  method}

The hypercubic blocking (HYP) method \cite{Hasenfratz:2001} as well as the APE smearing method \cite{albanese87} is frequently used to reduce the statistical errors.
Although this method is frequently used as a noise reduction, it should be remarked that HYP modifies the value of the static potential in the short distance $r/a<2$ where $r$ is the distance measured on the lattice and $a$ is the lattice spacing.
Therefore, we have applied HYP only to the $8 \times 8$ Wilson loop operator, while the original variable is used for the plaquette variable $U_P$ in the correlation function (\ref{cf1-1}). 

More concretely, the HYP renews the link variable $U_{x,\mu}$ as follows. 
First, we construct the decorated links $\tilde{U}_{x,\mu;\nu}$ from the original links $U_{x,\mu}$ by performing the 2 step projected APE smearing as follows. 
\begin{align} 
\tilde{U}_{x,\mu;\nu}  =& {\it Proj_{SU(2)}}
\left[
 (1-\alpha_2)U_{x,\mu}+\frac{\alpha_2}{4}\sum_{\pm\rho(\ne \nu,\mu)}
\bar{U}_{x,\rho;\nu \mu}\bar{U}_{x+\hat{\rho},\mu;\rho\nu}\bar{U}^{\dagger}_{x+\hat{\mu},\rho;\nu\mu}
\right] 
,
 \\
\bar{U}_{x,\mu;\nu\rho}  =& {\it Proj_{SU(2)}}
\left[ 
 (1-\alpha_3)U_{x,\mu}+\frac{\alpha_3}{2}\sum_{\pm\eta(\ne \rho,\nu,\mu)}
U_{x,\eta}U_{x+\hat{\eta},\mu}U^{\dagger}_{x+\hat{\mu},\eta}
\right] .
\label{cf3-2}
\end{align}
Second, we replace the link variable $U_{x,\mu}$ at site $x$ in the direction $\mu$  by the new link variable $U'_{x,\mu}$ which is obtained by the projected APE smearing from a set of the  decorated links $\tilde{U}_{x,\mu;\nu}$:
\begin{align} 
U_{x,\mu} \to U'_{x,\mu}= {\it Proj_{SU(2)}}
\left[
(1-\alpha_1)U_{x,\mu}+\frac{\alpha_1}{6}\sum_{\pm\nu(\ne \mu)}
\tilde{U}_{x,\nu;\mu}\tilde{U}_{x+\hat{\nu},\mu;\nu}\tilde{U}^{\dagger}_{x+\hat{\mu},\nu;\mu}
\right] .
\label{cf3-1}
\end{align}
Here the parameters $\alpha_1$, $\alpha_2$, $\alpha_3$ are chosen according to \cite{Hasenfratz:2001} as 
\begin{align} 
\alpha_1=0.75, \quad \alpha_2=0.6, \quad \alpha_3=0.3 .
\label{cf3-3}
\end{align}

The HYP preserves the gauge transformation property of the link variable, since the new link variable is obtained by summing up the clumps adjacent to the specified link variable, as in the APE smearing.

\end{document}